\documentclass[arguments]{aastex63}

\newcommand{\sm}{$M_\odot$}
\newcommand{\sL}{$L_\odot$}

\newcommand{\iras}{IRAS 15398$-$3359}

\newcommand{\hhco}{H$_{2}$CO}
\newcommand{\hhcs}{H$_{2}$CS}
\newcommand{\co}{C$^{18}$O}
\newcommand{\cchoh}{C$_2$H$_5$OH}
\newcommand{\meta}{CH$_3$OH}

\newcommand{\mecn}{CH$_3$CN}

\newcommand{\cchcn}{C$_2$H$_5$CN}
\newcommand{\hcccn}{HC$_3$N}
\newcommand{\nhhcho}{NH$_2$CHO}
\newcommand{\soo}{SO$_2$}
\newcommand{\kms}{km s$^{-1}$}
\newcommand{\mjybeam}{mJy beam$^{-1}$}
\received{}
\revised{}
\accepted{September 30, 2021}
\submitjournal{ApJ}

\shorttitle{}
\shortauthors{Okoda et al.}

\begin{document}

\title{Molecular Distributions of the Disk/Envelope System of L483: Principal Component Analysis for the Image Cube Data}

\correspondingauthor{Yuki Okoda}
\email{okoda@taurus.phys.s.u-tokyo.ac.jp}


\author{Yuki Okoda}
\affiliation{Department of Physics, The University of Tokyo, 7-3-1, Hongo, Bunkyo-ku, Tokyo 113-0033, Japan}

\author{Yoko Oya}
\affiliation{Department of Physics, The University of Tokyo, 7-3-1, Hongo, Bunkyo-ku, Tokyo 113-0033, Japan}
\affiliation{Research Center for the Early Universe, The University of Tokyo, 7-3-1, Hongo, Bunkyo-ku, Tokyo 113-0033, Japan} 
\author{Shotaro Abe}
\affiliation{Institute for Cosmic Ray Research (ICRR), The University of Tokyo, 5-1-5, Kahiwanoha, Kashiwa-shi, Chiba 277-8582, Japan}

\author{Ayano Komaki}
\affiliation{Department of Physics, The University of Tokyo, 7-3-1, Hongo, Bunkyo-ku, Tokyo 113-0033, Japan}

\author{Yoshimasa Watanabe}
\affiliation{Materials Science and Engineering, College of Engineering, Shibaura Institute of Technology, 3-7-5 Toyosu, Koto-ku, Tokyo 135-8548, Japan}

\author{Satoshi Yamamoto}
\affiliation{Department of Physics, The University of Tokyo, 7-3-1, Hongo, Bunkyo-ku, Tokyo 113-0033, Japan}
\affiliation{Research Center for the Early Universe, The University of Tokyo, 7-3-1, Hongo, Bunkyo-ku, Tokyo 113-0033, Japan} 


\begin{abstract}
\par Unbiased understandings of molecular distributions in a disk/envelope system of a low-mass protostellar source are crucial for investigating physical and chemical evolution processes.
We have observed 23 molecular lines toward the Class 0 protostellar source L483 with ALMA and have performed principal component analysis (PCA) for their cube data (PCA-3D) to characterize their distributions and velocity structures in the vicinity of the protostar.
The sum of the contributions of the first three components is 63.1\%.
Most oxygen-bearing complex-organic-molecule lines have a large correlation with the first principal component (PC1), representing the overall structure of the disk/envelope system around the protostar.
Contrary, the \co\ and SiO emissions show small and negative correlations with PC1.
The \nhhcho\ lines stand out conspicuously at the second principal component (PC2), revealing more compact distribution.
The HNCO lines and the high excitation line of \meta\ have a similar trend for PC2 to \nhhcho.
On the other hand, \co\ is well correlated with the third principal component (PC3).
Thus, PCA-3D enables us to elucidate the similarities and the differences of the distributions and the velocity structures among molecular lines simultaneously, so that the chemical differentiation between the oxygen-bearing complex organic molecules and the nitrogen-bearing ones is revealed in this source.
We have also conducted PCA for the moment 0 maps (PCA-2D) and that for the spectral line profiles (PCA-1D).
While they can extract part of characteristics of the molecular-line data, PCA-3D is essential for comprehensive understandings.
Characteristic features of the molecular-line distributions are discussed on \nhhcho.

\end{abstract}



\section{Introduction}
\label{intro}
\par 
Technological developments of radio astronomy bring on an increasing number of the detected molecular species, which enables us to obtain rich information on chemical composition and physical structure of astronomical objects.
In particular, Complex Organic Molecules (COMs), which consist of more than six atoms with one carbon atom at least, are found to be widely present in various environments of star formation: quiescent dense clouds, prestellar cores \citep[e.g.,][]{Bacmann et al.(2012), Cernicharo et al.(2012), Ceccarelli et al.(2017), Soma et al.(2018), ScibelliShirley(2020)}, disk/envelope systems of protostellar cores \citep[e.g.,][]{Cazaux et al.(2003), Bottinelli et al.(2004)b, Kuan et al.(2004), Pineda et al.(2012), Jorgensen et al.(2016), Oya et al.(2016), Lee et al.(2019), Imai et al.(2019), Bianchi et al.(2020)}, and outflow shock regions around protostars \citep[e.g.,][]{Arce et al.(2008), Sugimura et al.(2011), Codella et al.(2020), De Simone et al.(2020)}.
Many observations of COMs toward various sources reveal a chemical differentiation between nitrogen-bearing and oxygen-bearing COMs \citep[e.g.,][]{Wyrowski et al.(1999),  Bottinelli et al.(2004)b, Kuan et al.(2004), Beuther et al.(2005), Fontani et al.(2007), Calcutt et al.(2018), Oya et al.(2018), Csengeri et al.(2019)}.
These results indicate increasing complexity of chemical evolution in interstellar clouds and star forming regions.
In addition, nitrogen-bearing COMs generally tend to have a compact distribution in contrast to a rather extended distribution of oxygen-bearing COMs in star forming regions. 
Although these observed features are thought to be caused by some differences in the initial chemical composition of grain mantles before the birth of the protostar \citep{Charnley et al.(1992), Caselli et al.(1993)}, concrete processes responsible for the differentiation have not yet been unveiled.
To overcome this situation, more observations and analyses revealing the similarities and the differences of the molecular distributions are awaited for various sources.
It is, however, practically difficult to characterize many molecular-line data including velocity structures by eye, particularly in sensitive observations for a broad instantaneous bandwidth with various radio telescopes including Atacama Large Millimeter/submillimeter Array (ALMA) toward sources rich in molecular lines \citep*[e.g.,][]{Jorgensen et al.(2016), Imai et al.(2016), Watanabe et al.(2017)}. 
\par Principal Component Analysis (PCA) \citep{Jolliffe(1986)} is a useful tool for extracting even a small systematic difference of molecular line distributions around a protostar without any preconception.
\cite{Okoda et al.(2020)} conducted the PCA for the molecular line distributions (i.e., integrated intensity distribution) of the Class 0 protostellar source, \iras, and revealed that it can indeed extract the characteristic features of the molecular line distributions.
PCA has also been applied to the molecular line distributions successfully for analyses of radio astronomical observation data for external galaxies, galactic molecular clouds \citep*[e.g.,][]{Ungerechts et al.(1997), Meier & Turner(2005), Watanabe et al.(2016)}, and a starless core \citep{Spezzano et al.(2017)}.
If we analyze not only the distributions but also the velocity structures around protostars simultaneously, we can fully characterize the molecular line distribution by taking its kinematic information into account.
In this case, we can make a full use of huge observation data now available with the state-of-the-art interferometers and single-dish telescopes.
This is particularly important for studies on the disk/envelope systems of protostar sources, which generally have a complex velocity structure.
Hence, we here examine the application of PCA for image cube data around a protostar at a 100 au scale.
For this purpose, L483 is selected as the target, because this source is rich in various COMs in the vicinity of the protostar \citep{Oya et al.(2017), Agundez et al.(2019), Jacobsen et al.(2019)}.
As far as we know, there is no previous report on the PCA for the cube data of any protostellar sources.

\par L483 is a dark cloud in the Aquila Rift harboring the Class 0 protostellar source IRAS 18148$-$0440 \citep{Fuller et al.(1995), Chapman et al.(2013)}, whose bolometric luminosity is 10$-$14 \sL\ \citep{Ladd et al.(1991), Shirley et al.(2000), Tafalla et al.(2000)}. 
The distance to L483 is still controversial.
The distance to Aquila has been revised to be 436$\pm$9 pc by a recent survey \citep{Ortiz-Leon et al.(2018)}.
However, the Gaia-DR2 catalog within 1\degr\ of L483 shows that its location is closer, as supported by \cite{Jacobsen et al.(2019)}.
Therefore, the conventional distance to L483 of 200 pc \citep{DameThaddeus(1985)} is adopted in this paper.
Previous single-dish observations suggested that L483 is a possible candidate for Warm Carbon-Chain Chemistry (WCCC) source because of the relatively high abundance of C$_4$H \citep{Sakai et al.(2009), Hirota et al.(2009), Hirota et al.(2010), SakaiYamamoto(2013)}. 
It is also known as an interesting source for astrochemistry, where some peculiar molecules such as HCCO, HCS, HSC, and CNCN have been detected for the first time \citep*[e.g.,][]{Agundez et al.(2015a),Agundez et al.(2015b),  Agundez et al.(2018), Agundez et al.(2019)}.
On the other hand, \cite{Oya et al.(2017)} detected the compact emission of COMs in the vicinity of the protostar on a 100 au scale based on the ALMA observation, which reveals characteristic of hot corinos.
Hence, L483 shows a hybrid character of WCCC and hot corino chemistry at different scales.
\cite{Jacobsen et al.(2019)} also reported the detection of various COM lines in this source at a higher angular resolution (0\farcs1$-$0\farcs3; 20 au$-$60 au).
\par Many works have also been done for physical structures of L483.
Particularly, structures of the outflow extending along the east to west axis and the disk/envelope system have extensively been studied since 1990s. \citep*[e.g.,][]{Fuller et al.(1995), Park et al.(2000), Tafalla et al.(2000), Jorgensen(2004), Takakuwa et al.(2007), Oya et al.(2017), Oya et al.(2018), Jacobsen et al.(2019)}.
\cite{Park et al.(2000)} evaluated the outflow position angle to be 95\degr\ based on the HCO$^+$ ($J=1-$0) observation. 
\cite{Chapman et al.(2013)} later suggested the position angle of a magnetic pseudodisk to be 36\degr\ based on the 4.5 $\mu$m Spitzer image, as well as the outflow position angle of 105\degr\ based on the shocked H$_2$ emission reported by \cite{Fuller et al.(1995)}.
The kinematics of the disk/envelope system on a few 100 au scale or smaller has been studied by using molecular lines observed with ALMA.
\cite{Oya et al.(2017)} analyzed the velocity structure of the CS ($J=5-$4) emission by the infalling-rotating model and roughly evaluated the protostellar mass and the radius of the centrifugal barrier to be 0.1$-$0.2 \sm\ and 30$-$200 au, respectively.
They also suggested that the CS ($J=5-$4) emission partly traces the Keplerian disk within the centrifugal barrier.
Meanwhile, \cite{Jacobsen et al.(2019)} favoured an infall motion around the protostar on the basis of kinematics of the CS ($J=7-$6) and H$^{13}$CN ($J=4-$3) emission observed at a higher resolution ($\sim$40 au), which means the absence of a Keplerian disk down to at least 15 au in radius.
The outflow inclination angle was evaluated to be between 75\degr\ and 90\degr\ (0\degr\ for a pole-on configuration) with the ALMA observations of the CS and CCH line emission by \cite{Oya et al.(2018)}, which confirms the nearly edge-on configuration of the disk/envelope system. 
Thus, high angular resolution observations with ALMA have allowed us to investigate the disk/envelope system in detail as well as the outflow structure.
\par The disk/envelope system of L483 harbors various COMs, as mentioned above.
Although their distributions are concentrated around the protostar, the spectra toward the protostar position as well as the position-velocity diagram along the disk/envelope midplane show significant differences \citep{Oya et al.(2017), Oya et al.(2018), Jacobsen et al.(2019)}.
This implies the differentiation of the COM distributions.
It is of fundamental importance to elucidate the characteristics of the distributions of COMs in terms of the physical structure for exploring their production chemistry through their comparisons with the results of the other sources.
With this in mind, we apply PCA to the observed cube data in order to investigate the molecular distributions in L483 in detail.


\section{Observation}
\par Two ALMA observations toward L483 were carried out in the Cycle 4 operation on 2017 July 8. 
Spectral lines of SO (N$_J=5_6-4_5$), \meta\ (10$_{3,7}-11_{2,9}$, E; 4$_{2,3}-3_{1,2}$, E; and 18$_{3,16}-17_{4,13}$, A), \mecn, \hhco\ ($3_{0,3}-2_{0,2}$ and $3_{2,2}-2_{2,1}$), HNCO (10$_{0,10}-9_{0,9}$), \hhcs, \hcccn, and \co\ listed in Table \ref{observations} were observed in the frequency range from 218$-$237 GHz. 
For this observation, 40 antennas were used, where the baseline length ranged from 16 to 2647 m.
The total on-source time was 14.48 minutes.
The primary beam (half-power beam) width was 24\farcs6.
Spectral lines of \meta\ (4$_{2,2}-5_{1,5}$, A and 5$_{1,4}-4_{1,3}$, A), SO (N$_J=6_7-5_6$), \soo, CS, HNCO (12$_{0,12}-11_{0,11}$), \cchcn, \nhhcho, \cchoh, and SiO also listed in Table \ref{observations} were observed in the frequency range from 243$-$264 GHz.
For this observation, 42 antennas were used, where the baseline length ranged from 16 to 2647 m.
The total on-source time was 16.52 minutes, and the primary beam (half-power beam) width was 23\farcs8.
Note that, in Table \ref{observations}, the molecular lines are listed in a decreasing order of the first principal component derived later in the PCA for the cube data for easy comparison with the PCA results.
\par Both observations were conducted with the Band 6 receiver.
The backend correlator for molecular line observations except for \nhhcho\ (12$_{0,12}-11_{0,11}$) was set to a resolution of 122 kHz and a bandwidth of 59 MHz, and that for the \nhhcho\ (12$_{0,12}-11_{0,11}$) observation was set to a resolution of 282 kHz and a bandwidth of 234 MHz.
The field centers were taken to be ($\alpha_{2000}$, $\delta_{2000}$)= (18\fh17\fm29\fs947, $-$04\arcdeg39\arcmin39\farcs55).
The bandpass calibrator and the pointing calibrator were J1751+0939.
The flux calibrator and the phase calibrator were J1733-1304 and J1743-0350, respectively.
\par The data were reduced by Common Astronomy Software Applications package (CASA) 5.4.1
\citep{McMullin et al.(2007)} using a modified version of the ALMA calibration pipeline.
Phase self-calibration was performed using the continuum data, and then the solutions were applied to the spectral line data. After the self-calibration procedures, the data images were prepared by using the CLEAN algorithm, where the Briggs' weighting with a robustness parameter of 0.5 was employed.
The largest angular size is 2\farcs0 for both of these observations.
The original synthesized beam sizes are summarized in Table \ref{observations}.

\section{Data for PCA}
\label{sec-data}
\par  In this paper, we employ 23 molecular lines listed in Table \ref{observations}.
Since L483 is fairly rich in molecular lines, we only use the lines without apparent contamination from other lines.
Here, line contamination is checked by using the spectral line databases: CDMS \citep{Endres et al.(2016)} and JPL \citep{Pickett et al.(1998)}.
We focus on the distributions in the 2$\farcs0\times2\farcs0$ area around the protostar which covers the disk/envelope system, and the kinematics in the velocity range from -2.9 \kms\ to 14.25 \kms\ \citep[$V_{\rm sys}\sim$5.5 \kms;][]{Hirota
et al.(2009)}.
The latter is based on the spatial line features observed toward the protostar position \citep{Oya et al.(2017)}.
The spatial area and the velocity range correspond to 80$\times$80 pixels and 50 velocity channels, respectively.
To compare the molecular-line data at the same spatial and velocity resolution, the beam size is set to be 0$\farcs3\times0\farcs$3 with a Gaussian kernel by using $imsmooth$, which is the task of the CASA \citep{McMullin et al.(2007)}, and the velocity resolution is to be 0.35 \kms.
The moment 0 maps and the spectral line profiles toward the continuum peak are prepared by using the 0$\farcs3\times0\farcs3$ resolution cube, as shown in Figures \ref{moment_mol} and \ref{spectral_mol}, respectively.
Distributions of the most molecular lines are concentrated around the protostar.
Notable exceptions are the \co\ and SiO lines.
These peak positions are apparently shifted from the protostar toward the northeastern direction. 
Therefore, they can be regarded as an outlier in its following analysis.
\par In Section \ref{sec-3d}, we present the results of a PCA for the cube data (PCA-3D) to characterize the molecular distributions.
For reference, we also conduct a PCA for the moment 0 maps (PCA-2D; Section \ref{sec-2d}) and that for the spectral line profiles (PCA-1D; Section \ref{sec-1d}).
Three times the rms noise level is employed as the threshold level for the PCAs (Table \ref{observations}).
\par Since the correlation matrix is used to find the principal axes, the average of the original data is shifted to zero, and the distribution is normalized by the variance. Hence, the principal components (PCs) {\bf y} are written as:
\begin{equation}
{\bf y}={\bf Ax^*},
\end{equation}
where {\bf A} is the transformation matrix for the diagonalization of the correlation matrix and {\bf x$^*$} is the normalized distribution. The $j$th vector (molecule) of {\bf x$^*$} is related to the original distribution as:
\begin{equation}
{\bf x^*}_j=({\bf x}_j-\overline{{\bf x}_j})/\sigma_j^{1/2},
\end{equation}
where $\overline{{\bf x}_j}$ and $\sigma_j$ are the average and the variance of the $j$th vector of the original distribution ${\bf x}_j$, respectively. Here, the second term of Equation (2), ${\overline{{\bf x}_j}}/{\sigma_j^{1/2}}$, just provides the offset on the distribution, and does not affect the distribution of PCs. Hence, we omit this term in the presentation of PCs in this paper. This allows us to display PCs compatible with the images of the original distribution.

\section{PCA for the Cube Data (PCA-3D)}
\label{sec-3d}
\par 
We here conduct PCA-3D of the 23 molecular lines.
We first calculate the correlation coefficients among the cube data of the molecular lines, and then, derive principal components (PCs) by diagonalization of the correlation matrix.
The eigenvalues and eigenvectors of the first 7 PCs are given in Table \ref{percent_3d}.
As shown in Table \ref{percent_3d}, PC1 has the largest contribution ratio, 45.0 \%.
The contribution ratios of PC2 and PC3 are 11.4 \%\ and 6.7 \%, respectively, and hence, the sum of the three contribution ratios is 63.1 \%.
Figure \ref{screeplot} shows a scree plot of the contribution ratios for the principal components.
In Figure \ref{screeplot}(a), the contributions of PC$i$ ($i\geq$3) are relatively smaller than the first two components.
Nevertheless, PC3 shows a characteristic distribution, as described later.
Hence, we consider the three components here.

\subsection{Characteristic features of the Principal Components}
\label{sec-3d-1}
\par
Figures \ref{result3d}(d), (e), and (f) show the velocity channel maps of the three principal components.
As shown in Figure \ref{result3d}(d), PC1 represents a distribution around the protostar.
This is clearly seen in the moment 0 map of PC1 (Figure \ref{result3d}(a)).
In particular, this component (PC1) slightly extends toward the north and south directions at the blueshifted and redshifted velocities, respectively (Figure \ref{result3d}(d)).
These structures represent a rotation motion of the disk/envelope system, as reported previously \citep{Oya et al.(2017), Jacobsen et al.(2019)}.
Figure \ref{spectral_pc} shows the spectral line profiles of the principal components toward the continuum peak,
which are prepared for a circular region with a diameter of 0\farcs5 centered at the the continuum peak position.
The spectral line profile of PC1 shows a broad line width with an intensity dip around the systemic velocity, 5.5 \kms\ (Figure \ref{spectral_pc}(a)).
Moreover, the peak intensity of the redshifted component is stronger than that of the blueshifted one.
A similar trend of the spectrum is reported for a few molecular lines such as CS ($J=5-$4 and $7-6$), SO ($J_N=6_7-5_6$), HCN ($J=4-$3), and HCO$^+$ ($J=4-$3) \citep{Oya et al.(2017), Jacobsen et al.(2019)}. 
\par PC2 represents a compact distribution around the protostar, as seen in the moment 0 map of PC2 (Figure \ref{result3d}(b)).
It is more compact than that of PC1.
Although the channel maps of PC2 (Figure \ref{result3d}(e)) are noisy because of a relatively low contribution, the following features are seen: its distribution mostly shows faint or even negative intensities around the systemic velocity (5.5 \kms), whereas the emission toward the protostar is seen in the maps of 9.35 \kms\ and 11.1 \kms\ (Figure \ref{result3d}(e)).
This is also shown up in the spectral line profile of PC2 (Figure \ref{spectral_pc}(b)).
The two intensity peaks are visible around 0 \kms\ and 10 \kms, where the redshifted one is much stronger than the blueshifted one.
These peaks are offset by $\sim\pm$ 5 \kms\ from the systemic velocity.
The blueshifted one is lower than 3$\sigma$ lebel of PC2, so that it cannot be seen clearly in the channel maps.
Thus, PC2 reproduces the distribution concentrated around the protostar, which reveals the high velocity components.
\par
Figures \ref{result3d}(c) and \ref{result3d}(f) show the moment 0 map and the velocity channel maps of PC3, respectively.
The moment 0 map shows the positive distribution extending to northeast from the protostar and the negative crescent-like distribution surrounding the protostar on its southern side. In Figure \ref{result3d}(f), the former is marginally seen in the 0.6 \kms\ panel, whereas the latter is recognized in the 5.85 and 7.6 \kms\ panels. Figure \ref{spectral_pc}(c) shows the spectrum of PC3 for the 0\farcs5 area around the protostar. It has the intensity peaks at -1 \kms\ and 4 \kms, both of which are seen in the -1.15 \kms\ and 4.1 \kms\ panels of Figure \ref{result3d}(f). 
A broad dip around 8 \kms\ in the PC3 spectrum is also consistent with the channel maps. In short, PC3 represents rather extended positive and negative distributions in the northeastern and southern parts, respectively. 

\subsection{Characteristics of the molecular lines extracted by the PCA}
\label{sec-3d-mol}
\par In order to investigate the molecular-line distributions, we calculate the correlation coefficients between the principal components and the cube data for each molecular line, by using the method described by \cite{Okoda et al.(2020)} (See also Appendix \ref{SD_PCA}).
A large correlation coefficient means that the corresponding principal component well represents the distribution and the velocity structure of the corresponding molecular line.
Figure \ref{loading_3d}(a) shows the correlation coefficients for PC1, where the uncertainties due to the observation noise are evaluated by the  method described in Appendix \ref{SD_PCA}.
The 16 molecular lines have the correlation coefficients larger than 0.5, and hence, these molecular-line data can mostly be reproduced by using PC1: these lines indeed have the characteristic distribution and spectral profile of PC1 described above (Figures \ref{moment_mol} and \ref{spectral_mol}).
On the other hand, \co\ and SiO have a small negative correlation coefficient of PC1 (Figure \ref{loading_3d}(a)).

\par
The molecular lines less correlated with PC1 (lines 17$-$23) tend to have relatively large correlation coefficients of PC2, PC3, or both of them (Figures \ref{loading_3d}(b) and (c)).
Figure \ref{loading_3d}(b) shows that the two \nhhcho\ (12$_{11,1}-11_{11,0}$ and 12$_{0,12}-11_{0,11}$) lines have the largest positive correlation with PC2.
These characteristic features are discussed later (Section \ref{sec-disnh}).
The HNCO (10$_{0,10}-9_{0,9}$ and 12$_{0,12}-11_{0,11}$) lines and the high excitation line of \meta\ (18$_{3,16}-17_{4,13}$, A) have the similar trend to \nhhcho.
On the other hand, \co\ has the negative correlation coefficient with PC2 as large as -0.6.
It has the largest negative correlation coefficient (-0.75) with PC3, and hence, PC3 largely contributes to the distribution of \co.
In fact, the negative "crescent" part in the moment 0 map of PC3 (Figure \ref{result3d}(c)) resembles the \co\ distribution (Figure \ref{moment_mol}(v)).
\cchcn\ and SiO are positively correlated with PC3, which can reproduce an intensity dip around the systemic velocity and faint high velocity components in each line profile (Figures \ref{spectral_mol}(q) and (w)).
Furthermore, both SiO and PC3 indeed show the component elongated from the continuum peak to the northeastern direction (Figure \ref{moment_mol}(w) and Figure \ref{result3d}(c)).

\subsection{Velocity Structure of PC1}
\par Since most molecular lines are correlated with PC1, we prepare the moment 1 map of PC1 (Figure \ref{moment1_pc1}(a)) in order to study the $'$standard$'$ velocity structure of our data set.
The velocity gradient can clearly be seen along the disk/envelope direction (P.A. 15\degr), as reported previously \citep{Oya et al.(2017)}.
The position velocity (PV) diagrams of PC1 along the disk/envelope direction and the direction perpendicular to it are shown in Figures \ref{moment1_pc1}(b) and (c), respectively.
The PV diagram along the disk/envelope direction shows a clear velocity gradient (Figure \ref{moment1_pc1}(b)).
On the other hand, an apparent gradient cannot be seen in the PV diagram along the line perpendicular to the disk/envelope system (Figure \ref{moment1_pc1}(c)).
Hence, the velocity structure of PC1 is likely a rotation motion around the protostar.
\par This result is consistent with the previous reports.
\cite{Oya et al.(2017)} interpreted the velocity structure of CS ($J=5-$4) as a combination of the infalling rotating motion and the Keplerian motion around the protostar.
They also reported that the velocity gradient of SO ($J_N=6_7-5_6$) originates from the rotation motion in the inner part of the disk/envelope system.
Although these molecular distributions are different from each other on a scale larger than 2\arcsec, as reported by \cite{Oya et al.(2017)}, they are essentially similar to each other on a smaller scale considered in this study, as revealed by high correlations of CS ($J=5-$4) and SO ($J_N=6_7-5_6$) with PC1.
The kinematic structure of CS ($J=7-$6) line was also investigated by \cite{Jacobsen et al.(2019)}, whose moment 1 map shows the velocity gradient along the north to south axis.
They further reported the rotation motion by using the lines of complex organic molecules (COMs).
Thus, PCA-3D successfully extracts not only the most common distribution but the associated velocity structure.
Although the lines showing the high correlation with PC1 (lines 1-16) come from the disk/envelope system around the protostar, it is difficult to distinguish between the infalling rotating motion and the Keplerian motion from this result.
\par It should be noted that PC2 and PC3 have negative intensities in their channel maps (Figures \ref{result3d}(e) and (f)). 
For these cases, moment 1 maps are not very useful.
We need to directly inspect the principal component data (velocity channel maps) to discuss their velocity structures.

\section{PCA for the moment 0 maps (PCA-2D)}
\label{sec-2d}
\par In this section, we apply the PCA for the 23 molecular-line images integrated over the velocity axis of the cube data used in Section \ref{sec-3d} (moment 0 maps: Figure \ref{moment_mol}), and compare the results to those of PCA-3D.
The result of PCA-2D is shown in Table \ref{percent_2d}.
The contribution ratio of PC1 is 77.9 \%, and hence, molecular-line distributions can mostly be represented by PC1.
The contribution ratios of PC2 and PC3 are 8.1 \% and 5.5 \% (Table \ref{percent_2d} and Figure \ref{screeplot}(b)).
We here discuss the first three components, the sum of whose ratios is 91.5 \%.

\par Figures \ref{result_2d}(a), (b), and (c) show the maps of PC1, PC2, and PC3 of PCA-2D, respectively.
PC1 has a distribution concentrated around the continuum peak (Figure \ref{result_2d}(a)), which looks similar to the moment 0 map of PC1 of PCA-3D, while its contribution ratio is much larger than that in PCA-3D (Figure \ref{result3d}(a)).
PC2 has an extended distribution surrounding the continuum peak from the north to the southeast (Figure \ref{result_2d}(b)). 
PC3 extends from the vicinity of the continuum peak to the northeastern direction (Figure \ref{result_2d}(c)), most of which is concentrated in the northeastern part. 

\par According to Figure \ref{result_2d}(d), PC1 has the correlation coefficients larger than 0.5 for all molecular lines except for \co\ and SiO.
The large correlation coefficients for most molecular lines are consistent with their large contribution ratios of PC1.
In Figures \ref{result_2d}(e) and (f), the two molecular lines, \co\ and SiO, are well correlated with PC2 and PC3, respectively, which means that these molecular lines have peculiar distributions among the 23 molecular lines.
PC2 and PC3 (Figures \ref{result_2d}(b) and (c)) look similar to the moment 0 maps of \co\ and SiO, respectively (Figures \ref{moment_mol}(v) and (w)).
This feature is also seen in the PC1-PC2, PC1-PC3, and PC2-PC3 {biplots} (Figures \ref{result_2d_biplot}(a), (b), and (c)), which display the contributions of the principal components for each molecular-line distribution.
In these figures, all molecular lines except for the two species have a relatively small contribution of PC2 and PC3.
Slightly positive and negative PC2 values could indicate the extended and compact distributions, respectively. 
For example, SO (\#3 and \#5) and CS (\#12), which have small positive values of PC2, tend to have an extended distribution. On the other hand, small negative values of PC2 for the two lines of \nhhcho\ (Figure \ref{result_2d_biplot}(a): \#19 and \#21) may indicate their compact distribution, as seen in PCA-3D.
\co\ and SiO are located at the large positive parts of PC2 and PC3 on the PC2-PC3 biplot, respectively (Figure \ref{result_2d_biplot}(c)).
Thus, PC3 almost reproduces the distribution of SiO extending from the continuum peak to the northeastern direction (Figure \ref{moment_mol}(w)).
Such a distribution of SiO was also reported by \cite{Oya et al.(2017)}.
Since SiO is known as a shock tracer, the characteristic distribution may represent a shock related to the outflow: outflows launched from the protostar would likely hit a gas clump remaining around the protostar.
On the other hand, \hcccn\ has a positive value of PC3 larger than the others, indicating that its distribution shows some similarity to that of SiO: its peak position is slightly shifted toward the northeast from the continuum peak.
However, it seems early to conclude that these two species are chemically related.
\par It should be noted that the peculiar distributions of \co\ and SiO can also be identified by PCA-3D.
On the contrary, the molecular lines having a rather broad component such as \nhhcho\ are not clearly identified in the PCA-2D.



\section{PCA for the spectral line profiles (PCA-1D)}
\label{sec-1d}
\par 
In addition to the PCAs described above, we apply PCA for the spectral line profiles toward the continuum peak (Figure \ref{spectral_mol}) in the same velocity range as the cube data.
The spectra are prepared for a circular region with a diameter of 0\farcs5 centered at the continuum peak to fairly consider the molecular lines extending around the protostar.
The signal-to-noise ratios of the \cchcn, \co, and SiO spectra are low toward the continuum peak (Figures \ref{spectral_mol}(q), (v) and (w)).
As well, the \nhhcho\ (12$_{1,11}-11_{1,10}$ and 12$_{0,12}-11_{0,11}$) lines are not bright enough for PCA-1D because of their compact distributions (Figures \ref{moment_mol}(s) and (u)).
Hence, we here conduct PCA-1D for the18 molecular line except for these five lines.
The result of PCA-1D is shown in Table \ref{percent_1d}.
As seen in Table \ref{percent_1d}, PC1 has the largest contribution ratio, 64.1 \%.
The contribution ratios of PC2 and PC3 are 14.1 \% and 7.1 \%, respectively, while that of PC4 is 4.1 \%.
This result is also seen in Figure \ref{screeplot}(c).
We here discuss the first three components, the sum of whose contributions is 85.3 \%.

\par 
Figures \ref{result_1d_a}(a), (b), and (c) show the PC1, PC2, and PC3 of PCA-1D, respectively.
PC1 look quite similar to the spectral line profile of PC1 obtained for PCA-3D.
On the other hand, PC2 and PC3 of PCA-1D are different from those of PCA-3D.
 In PCA-1D, PC2 shows the intensity peak around the systemic velocity  (5.5 \kms) with a blueshifted shoulder.
PC3 has the intensity dips around 3 \kms\ and 10 \kms.
The lack of the data of \cchcn, \co, SiO, and \nhhcho\ (12$_{0,12}-11_{0,11}$ and 12$_{1,11}-11_{1,10}$) in PCA-1D can be responsible for the differences of PC2 and PC3 between PCA-3D and PCA-1D.
\par As shown in Figure \ref{result_1d_a}(d), most molecular lines show the large correlation coefficients of PC1.
Molecular lines less correlated with PC1 tend to have the large correlation coefficients with PC2 or PC3 (Figures \ref{result_1d_a}(e) and (f)).
\hcccn (\#18), HNCO (12$_{0,12}-11_{0,11}$;\#15), \meta\ (18$_{3,16}-17_{4,13}$, A; \#13), HNCO (10$_{0,10}-9_{0,9}$;\#14), and H$_2$CS (\#16) are well correlated with PC2 (Figures \ref{result_1d_a}(e)).
The positive and negative PC2 values mean strong and weak blueshifted components, respectively, of these lines.
On the other hand, \cchoh\ has the correlation coefficients with PC3 as large as 1.0 (Figure \ref{result_1d_a}(f)), and can almost be reproduced by PC3 with a little contribution of PC1.
This means that only this molecular line has the peculiar spectral line profile, although it may be affected by the low signal-to-noise ratio.
However, this feature should be confirmed by observing the other lines.
The characteristics of the spectral line profiles are thus extracted to some extent by PC2 and PC3, which are seen more clearly in the PC1-PC2, PC1-PC3, and PC2-PC3 biplots, respectively (Figures \ref{result_1d_b}). 
The results of PCA-1D are essentially consistent with the results of PCA-3D.
It should be noted that the result would depend on the selected area for the spectral line profiles.

\section{Discussion}
\label{sec-disnh}
\par We have conducted the three PCAs for the molecular-line data described above.
We can extract some characteristic features of the molecular lines even with PCA-2D and PCA-1D.
However, PCA-2D does not consider the velocity information, and PCA-1D loses the information on their spatial distributions.
For this reason, we need a caution in characterization of the molecular line distributions based on PCA-2D and PCA-1D.
PCA-3D is thus desirable for full identification of the similarities and differences of the molecular-line distributions.
In this section, we discuss the characteristic features of the molecular line distributions derived from the results of PCA-3D.

\subsection{Nitrogen-bearing and Oxygen-bearing Species}
\par Figures \ref{plot_3d}(a), (b), and (c) are biplots showing the contributions of the principal components for each molecular-line distribution derived from PCA-3D on the PC1-PC2, PC1-PC3, and PC2-PC3 planes, respectively.
The yellow and blue plots represent the nitrogen-bearing and oxygen-bearing organic molecules, respectively.
In Figures \ref{plot_3d}(a) and (b), the PC1 contributions of the oxygen-bearing molecules tend to be  slightly larger than those of the nitrogen-bearing molecules.
Moreover, the oxygen-bearing {\bf organic} molecules except for \meta\ (18$_{3,16}-17_{4,13}$, A) show small contributions of PC2 and PC3.
On the contrary, the contributions of PC2 and PC3 are larger for the nitrogen-bearing molecules except for CH$_3$CN than those for the oxygen-bearing molecules.
This result reveals a spatial differentiation of the nitrogen-bearing and oxygen-bearing molecular-line emission in the disk/envelope system of L483. 
In principle, distributions of molecular lines can be different due to the excitation effect.
Although we cannot see this trend in the lower excitation \meta\ lines ($E_u=$50: \#1, $E_u=$61: \#2, $E_u=$46 K: \#4, and $E_u=$190 K: \#6), only the very high excitation line of \meta\ (18$_{3,16}-17_{4,13}$, A; $E_u=$447 K: \#13) shows a relatively high contribution of PC2.
The dust temperature is supposed to be 70$-$130 K \citep{Oya et al.(2017), Jacobsen et al.(2019)}, whereas the H$_2$ density is roughly estimated to be 10$^8$ cm$^{-2}$ or higher from the H$_2$ column density (4.9$-$9.5)$\times10^{23}$ cm$^{-2}$ \citep{Oya et al.(2017)} and the assumed size of the hot region ($\sim$100 au).
Although the LTE condition is almost fulfilled, the line with high $E_u$ such as  \meta\ (18$_{3,16}-17_{4,13}$, A) is expected to appear in a compact hot region around the protostar.
This is the reason for the relatively high PC2 value.
\par Judging from the large contribution of PC1, the distributions of the oxygen-bearing molecules slightly extend along the disk/envelope direction.
On the other hand, the nitrogen-bearing molecules with the small PC1 and large PC2 contributions have the compact distributions with the high velocity components.
The different distributions are also seen in the PC3 values for some molecules, such as \cchcn\ and \hcccn\ (Figure \ref{plot_3d}(c)).
A larger velocity width for a more compact distribution found in PC2 is physically reasonable, because the rotation/infall velocity is expected to increase as approaching to the protostar.
\par As mentioned in Section \ref{intro}, the chemical differentiation between the nitrogen-bearing and oxygen-bearing complex organic molecules (COMs) have been reported for various star-forming regions \citep*[e.g.,][]{Wyrowski et al.(1999), Bottinelli et al.(2004)a, Bottinelli et al.(2004)b, Kuan et al.(2004), Beuther et al.(2005), Fontani et al.(2007), Calcutt et al.(2018), Oya et al.(2018), Csengeri et al.(2019)}.
Thus, this still remains as an important issue to be solved for astrochemistry.
\cite{Csengeri et al.(2019)} recently conducted the ALMA observation toward the high-mass protostellar source, G328.2551-0.5321, and reported that the nitrogen-bearing COMs including \nhhcho\ and \cchcn\ are concentrated around the protostar, while the oxygen-bearing COMs such as \meta\ and \cchoh\ are present around the centrifugal barrier of the infalling gas.
A similar situation for the oxygen-bearing COMs is also reported for IRAS 16293-2422 \citep{Oya et al.(2016)}.
\cite{Csengeri et al.(2019)} argued that the production of the nitrogen-bearing COMs is related to the protostellar heating, while that of the oxygen-bearing COMs is to the accretion shock heating of the infalling gas.
If this picture is applied to low-mass protostellar sources, the nitrogen-bearing COMs would have more compact distributions than the oxygen-bearing COMs.
This is consistent with our results of PCA-3D.
Thus, we reveal such a differentiation in the disk/envelope system of L483.
Note that the SiO emission also traces a shock in L483, but it is caused by the outflow.
Indeed, SiO does not have any correlation with the oxygen-bearing  molecules. 
It should be noted that \cchcn\ shows PC2 and PC3 similar to SiO (Figure \ref{plot_3d}(c)).
This similarity might tell us a clue to solve the origin of \cchcn.
To go further, we need to observe other lines of this molecule for confirmation of the above trend.
Although the number of the molecular lines involved in this study is limited, we can state that the PCA-3D can extract the chemical differentiation within the disk/envelope system in an unbiased way.
An extension of such studies for various sources is awaited.

\subsection{\nhhcho}
\label{sec-nhhcho}
\par 
An important result from this study is on the distribution of \nhhcho.
This molecule containing a peptide-like bond (N$-$C$=$O) is thought to be a mother molecule for prebiotic evolution, and hence, various observational, experimental, and theoretical works for the molecule have been reported \citep[e.g., Saladino et al. 2012;][]{Barone et al.(2015), Lopez-Sepulcre et al.(2015), Lopez-Sepulcre et al.(2019), Quenard et al.(2018), Martin-Domenech et al.(2020)}.
Detections of \nhhcho\ toward star-forming regions \citep[e.g.,][]{Bisschop et al.(2007), Yamaguchi et al.(2012), Kahane et al.(2013), Mendoza et al.(2014), Gorai et al.(2020)} as well as comets of the Solar System \citep{Bockelee-Morvan et al.(2000), Biver et al.(2014), Goesmann et al.(2015)} imply inheritance of this molecule from protostellar cores to planetary systems.
\par As described in Section \ref{sec-3d}, the \nhhcho\ (12$_{11,1}-11_{11,0}$ and 12$_{0,12}-11_{0,11}$) lines are well correlated with PC2 (Figure \ref{loading_3d}(b)).
This means that they have a compact distribution with the high velocity component.
A compact distribution of \nhhcho\ can also be confirmed in the map presented by \cite{Jacobsen et al.(2019)}.
Figures \ref{spectral_mol} (s) and (u) represent the line profiles of \nhhcho\ (12$_{11,1}-11_{11,0}$ and 12$_{0,12}-11_{0,11}$) toward the continuum peak, respectively.
They both have a slight dip at the systemic velocity and a redshifted peak around $\sim$9 \kms, as seen in the PC2 spectrum of PCA-3D, although the slight difference is seen between the two lines.
Such a spectral line feature of the \nhhcho\ (12$_{0,12}-11_{0,11}$) line is consistent with that reported in \cite{Oya et al.(2017)}.
The compact distribution of \nhhcho\ is not likely the excitation effect, because the upper state energies ($E_u$) of the \nhhcho\ lines are comparable to those of the low-excitation \meta\ lines.
Thus, we conclude that \nhhcho\ resides in the inner part of the rotating disk/envelope structure.
This result suggests that \nhhcho\ is liberated from dust grains or formed in the gas phase in the higher temperature region near the protostar than the oxygen-bearing COMs.
In the other low-mass protostellar source, B335, the \nhhcho\ (12$_{0,12}-11_{0,11}$) emission is also concentrated toward the protostar. 
It has a broad spectral line width without a strong absorption feature around the systemic velocity \citep{Imai et al.(2016), Imai et al.(2019)}.
These characteristic features of the \nhhcho\ line in B335 are similar to the case of L483.
\par Various chemical pathways for \nhhcho\ have been proposed including gas-phase \citep{QuanHerbst(2007), Garrod et al.(2008), Halfen et al.(2011), Barone et al.(2015), Skouteris et al.(2017)} and grain-surface processes \citep{Charnley(1997),Raunier et al.(2004), Fedoseev et al.(2016)}.
\cite{Mendoza et al.(2014)} and \cite{Lopez-Sepulcre et al.(2015)} reported a correlation between the abundance of \nhhcho\ and HNCO based on their survey observations toward various star-forming regions from low-mass to high-mass ones.
The abundance relationship was also reported with the ALMA observations \citep{Coutens et al.(2016), Allen et al.(2020)} and was supported by theoretical calculation \citep{Quenard et al.(2018)}.
These works suggest some chemical relations between the two species.
Hydrogenation of HNCO has been proposed as one possible formation process leading to \nhhcho\ on dust grains \citep{Lopez-Sepulcre et al.(2015), Quenard et al.(2018)}, although \cite{Quenard et al.(2018)} noted that the power-law correlation between the two species can be seen, even if \nhhcho\ is not formed from HNCO through hydrogenation on dust grains.
Thus, the chemical link to HNCO would be a key to resolve the formation route of \nhhcho.
In this study, a relation between HNCO and \nhhcho\ is also suggested in PCA-3D.
HNCO (10$_{0,10}-9_{0,9}$ and 12$_{0,12}-11_{0,11}$) are positively correlated with PC2 (Figure \ref{loading_3d}(b)) as the \nhhcho\ lines.
\par Nevertheless, we have to say that the formation process for \nhhcho\ is still controversial.
In general, the \nhhcho\ distribution seems to be more compact than O-bearing molecules in other sources: e.g., HH212 \citep{Lee et al.(2017)}, G328.2 \citep{Csengeri et al.(2019)}, G10.6 \citep{Law et al.(2021)}.
If we can extract the common characteristic distribution of \nhhcho\ in various sources with the PCA, we will have an important clue to its formation process.
To understand the chemical process in the disk/envelope system during the early stage of star formation, observations of more molecular lines for other Class 0/I protostars with higher angular resolution and higher signal-to-noise ratio are awaited.




\section{Summary}
\par 1. In order to fully characterize the distribution of molecular lines in the disk/envelope structure of the Class 0 protostellar source L483, we have conducted PCA-3D in the vicinity of the protostar.
We have used the 23 molecular lines without apparent contamination from other lines with ALMA.
The sum of the contributions of PC1, PC2, and PC3 is 63.1\%.
\par 2. In PCA-3D, PC1 represents the average distribution of the molecular-line data, and the 16 molecular lines are well correlated with PC1. 
We find that these lines trace a rotation motion of the disk/envelope system based on the PV diagram of PC1.
\co\ and SiO are notable exceptions, which have the small negative correlation of PC1.
\par 3. In PCA-3D, some molecular lines less correlated with PC1 tend to have large correlation coefficients of PC2. 
They are the \nhhcho\ (12$_{0,12}-11_{0,11}$ and 12$_{11,1}-11_{11,0}$), HNCO (10$_{0,10}-9_{0,9}$ and 12$_{0,12}-11_{0,11}$), and very high excitation \meta\ (18$_{3,16}-17_{4,13}$, A) lines, which have a compact distribution with high velocity components.
Association of the very high excitation \meta\ line with the compact region implies a hot and dense condition of the emitting region of \nhhcho\ and HNCO.
A similar trend of HNCO to \nhhcho\ suggests a possible chemical link between the two species.
 \par 4. The results of PCA-3D suggest a chemical differentiation of the nitrogen-bearing and oxygen-bearing molecular-line data in L483.
The oxygen-bearing molecules tend to have a distribution slightly more extended than the nitrogen-bearing molecules.
These features are mainly suggested by PC1 and PC2.
\par  5. Thus, PCA-3D successfully extracts the chemical differentiation within the disk/envelope system.
For reference, we have also performed PCA-2D and PCA-1D.
For these two PCAs, there is a limitation in extracting specific distribution features because of using the part of the data, and hence, special caution should be paid in interpretation of their results.
PCA-3D enables us the full characterization of the distribution and the kinematics of the various molecular lines simultaneously without any preconception.

\acknowledgments
We thank the anonymous reviewer for valuable comments.
This paper makes use of the following ALMA data set:
ADS/JAO.ALMA\# 2016.1.01325.S (PI: Yoko Oya). ALMA is a partnership of the ESO (representing its member states), the NSF (USA) and NINS (Japan), together with the NRC (Canada) and the NSC and ASIAA (Taiwan), in cooperation with the Republic of Chile.
The Joint ALMA Observatory is operated by the ESO, the AUI/NRAO, and the NAOJ. The authors thank to the ALMA staff for their excellent support.
This project is supported by a Grant-in-Aid from Japan Society for the Promotion of Science (KAKENHI: No. 18H05222, 19H05069, 19K14753, and 20J13783.
Yuki Okoda thanks the Advanced Leading Graduate Course for Photon Science (ALPS) and Japan Society for the Promotion of Science (JSPS) for financial support.

{}

\appendix

\section{Standard deviation of the PCA components}
\label{SD_PCA}
\par 
We derive the standard deviation for the principal component value of each molecular line \citep{Gratier et al.(2017), Spezzano et al.(2017), Okoda et al.(2020)} for each PCA.
As for the PCA-3D and PCA-2D,  we generate the Gaussian random noise for each image pixel (80$\times$80) and each velocity channel (only for the cube: 50 channel) of the molecular-line data.
The noise distribution is convolved by the beam size of the molecular lines (0$\farcs3\times0\farcs3$), where the noise level corresponds to 1$\sigma$ noise level of each image.
As for the PCA-1D, the Gaussian random noise is prepared for each channel of the molecular-line data.
We add these artificial noises to the original data for each molecular line and conduct PCA.
This procedure is repeated 1000 times, and we finally calculate the standard deviations of the PC values.
The standard deviations are shown by the grey dashed ellipses in the plots of the principal components (Figures \ref{result_2d_biplot}, \ref{result_1d_b}, and \ref{plot_3d}).
We also calculate the standard deviations for the correlation coefficients between the molecular lines and the principal components by using each eigenvalue and each eigenvector value as :
\begin{equation}
\delta{Cor ({\bf x^*}_j, {\bf y}_i)}=Cor ({\bf x^*}_j, {\bf y}_i){\sqrt{(\frac{\delta\lambda_i}{2\lambda_i})^2+(\frac{\delta z_{ji}}{z_{ji}})^2}},
\end{equation}
where $Cor ({\bf x^*}_j, {\bf y}_i)$ is the correlation coefficient defined as \citep{Jolliffe(1986)}:
\begin{equation}
Cor ({\bf x^*}_j, {\bf y}_i)={\sqrt{\lambda_i}z_{ji}},
\end{equation}
$\lambda_i$ stands for the eigenvalue for the $i$ th eigenvector for PC$i$ {\bf z$_{i}$}, and $z_{ji}$ does for the eigenvector component for the $j$ th molecular line.
$\delta\lambda_i$ and $\delta z_{ji}$ are the standard deviations of $\lambda_i$ and $z_{ji}$, respectively, which are derived by the method mentioned above.
The uncertainties thus derived are shown in grey in Figures \ref{loading_3d}, \ref{result_2d}(d-f), and \ref{result_1d_a}(d-f).


\begin{longrotatetable}
\begin{table}[ht]
\centering
\caption{Parameters of Observed Lines $^a$ \label{observations}}
\scalebox{0.8}{
\begin{tabular}{cccccccccc}
\hline \hline
 Number& Molecule&Transition & Frequency & $S \mu^2$ & $E_{\rm u}$$k^{-1}$ & Original beam size&Rms (Cube)&Rms (Moment 0)&Rms (Spectral)\\
& &&GHz&$D^2$&K&&\mjybeam &\mjybeam\kms &K\\
 \hline
1 & \meta & 5$_{1,4}-4_{1,3}$, A  & 243.915788 & 15.5 & 50 &  0\farcs228 $\times$ 0\farcs146 (P.A. -59$^{\circ}$)  & 4 & 14 & 0.5\\ 
2 & \meta & 4$_{2,2}-5_{1,5}$, A  & 247.228587 & 4.34 & 61 &  0\farcs227 $\times$ 0\farcs144 (P.A. -59$^{\circ}$)  & 4 & 14 &  0.5\\ 
3 & SO & 5$_6-4_5$  & 219.949442 & 14 & 35 &  0\farcs244 $\times$ 0\farcs141 (P.A. -65$^{\circ}$)  & 4 & 15 & 0.7\\ 
4 & \meta & 4$_{2,3}-3_{1,2}$, E  & 218.440063 & 13.9 & 46 &  0\farcs244 $\times$ 0\farcs142 (P.A. -65$^{\circ}$)  & 4 & 13 & 0.7\\ 
5 & SO & 6$_7-5_6$  & 261.843721 & 16.4 & 48 &  0\farcs241 $\times$ 0\farcs157 (P.A. -53$^{\circ}$)  & 4 & 15 & 0.5\\ 
6 & \meta & 10$_{3,7}-11_{2,9}$, E  & 232.945797 & 12.1 & 190 &  0\farcs232 $\times$ 0\farcs133 (P.A. -64$^{\circ}$)  & 4 & 14 & 0.6\\ 
7 & \hhco & 3$_{2,1}-2_{2,0}$  & 218.760066 & 9.06 & 68 &  0\farcs245 $\times$ 0\farcs142 (P.A. -65$^{\circ}$)  & 4 & 12 & 0.7\\ 
8 & \hhco & 3$_{2,2}-2_{2,1}$  & 218.475632 & 9.06 & 68 &  0\farcs245 $\times$ 0\farcs142 (P.A. -65$^{\circ}$)  & 4 & 14 & 0.7\\ 
9 & \soo & 14$_{0,14}-13_{1,13}$  & 244.2542183 & 28 & 94 &  0\farcs229 $\times$ 0\farcs145 (P.A. -59$^{\circ}$)  & 4 & 13 & 0.5 \\ 
10 & \mecn & 12$_{2,0}-11_{2,0}$  & 220.7302611 & 359 & 97 &  0\farcs242 $\times$ 0\farcs138 (P.A. -65$^{\circ}$)  & 4 & 12 & 0.7\\ 
11 & \hhco & 3$_{0,3}-2_{0,2}$  & 218.222192 & 16.3 & 21 &  0\farcs245 $\times$ 0\farcs142 (P.A. -65$^{\circ}$)  & 4 & 12 & 0.7\\ 
12 & CS & 5$-$4  & 244.9355565 & 19.2 & 35 &  0\farcs229 $\times$ 0\farcs146 (P.A. -58$^{\circ}$)  & 4 & 16 & 0.5\\ 
13 & \meta & 18$_{3,16}-17_{4,13}$, A  & 232.783446 & 21.8 & 447 &  0\farcs230 $\times$ 0\farcs132  (P.A. -64$^{\circ}$)  & 4 & 14 & 0.6\\ 
14 & HNCO & 10$_{0,10}-9_{0,9}$  & 219.798274 & 25 & 58 &  0\farcs245 $\times$ 0\farcs142 (P.A. -65$^{\circ}$)  & 4 & 12 & 0.7\\ 
15 & HNCO & 12$_{0,12}-11_{0,11}$  & 263.748625 & 30 & 82 &  0\farcs239 $\times$ 0\farcs157 (P.A. -53$^{\circ}$)  & 4 & 11 &  0.7\\ 
16 & \hhcs & 7$_{1,7}-6_{1,6}$  & 236.7270204 & 56 & 59 &  0\farcs229 $\times$ 0\farcs132 (P.A. -63$^{\circ}$)  & 4 & 12 & 0.6\\ 
17 & \cchcn & 11$_{4,8}-10_{3,7}$  & 260.5411331 & 6.96 & 46 &  0\farcs243 $\times$ 0\farcs158 (P.A. -53$^{\circ}$)  & 4 & 11 & 0.5\\ 
18 & \cchoh & 12$_{1,11}-11_{2,9}$  & 244.5874939 & 3.93 & 129 &  0\farcs229 $\times$ 0\farcs146 (P.A. -58$^{\circ}$)  & 4 & 13 & 0.5\\ 
19 & \nhhcho & 12$_{1,11}-11_{1,10}$  & 261.3274496 & 156 & 85 &  0\farcs241 $\times$ 0\farcs158 (P.A. -53$^{\circ}$)  & 4 & 12 & 0.5\\ 
20 & \hcccn & 26$-$25  & 236.5127888 & 362 & 153 &  0\farcs229 $\times$ 0\farcs132 (P.A. -63$^{\circ}$)  & 4 & 12 & 0.5\\ 
21 & \nhhcho & 12$_{0,12}-11_{0,11}$  & 247.390719 & 156 & 78 &  0\farcs227 $\times$ 0\farcs143 (P.A. -58$^{\circ}$)  & 4 & 13 & 0.5\\ 
22 & \co & 2$-$1  & 219.5603541 & 0.024 & 16 &  0\farcs245 $\times$ 0\farcs142 (P.A. -65$^{\circ}$)  & 4 & 12 & 0.7\\ 
23 & SiO & 6$-$5  & 260.518009 & 57.6 & 44 &  0\farcs243 $\times$ 0\farcs158 (P.A. -53$^{\circ}$)  & 4 & 14 & 0.5\\ 
\hline
\end{tabular}}
\begin{flushleft}
\tablecomments{$^a$ Line parameters are taken from CDMS \citep{Endres et al.(2016)}. 
The root-mean-square (rms) noise and the beam size are based on the observation data.
The molecular lines are ordered by PC1 of PCA-3D in Table \ref{percent_3d} for consistency in the numbering of molecular lines in this paper.
}
\end{flushleft}
\end{table}
\end{longrotatetable}

\begin{table}[ht]
\caption{Eigenvectors of the Principal Components and their Eigenvalues for PCA-3D\label{percent_3d}}
\scalebox{1.0}{
\begin{tabular}{ccccccccccccc}
\hline \hline
Number & Molecule & PC1 & PC2 & PC3 & PC4 & PC5 & PC6 & PC7 \\
\hline
1 & \meta\ (5$_{1,4}-4_{1,3}$, A) & 0.286 & -0.072 & 0.015 & 0.062 & -0.017 & -0.027 & -0.133 \\
2 & \meta\ (4$_{2,2}-5_{1,5}$, A)  & 0.278 & 0.003 & -0.019 & 0.021 & -0.055 & -0.075 & -0.113 \\
3 & SO\ (5$_6-4_5$) & 0.277 & -0.143 & -0.037 & -0.002 & -0.034 & 0.187 & 0.115 \\
4 & \meta\ (4$_{2,3}-3_{1,2}$, E)  & 0.276 & -0.041 & 0.084 & -0.099 & 0.043 & -0.002 & -0.13 \\
5 & SO\ (6$_7-5_6$)  & 0.27 & -0.142 & -0.138 & 0.051 & -0.073 & 0.181 & 0.122 \\
6 & \meta\ (10$_{3,7}-11_{2,9}$, E)  & 0.27 & -0.037 & 0.054 & 0.019 & 0.023 & -0.016 & 0.104 \\
7 & \hhco\ (3$_{2,1}-2_{2,0}$)  & 0.256 & -0.197 & 0.056 & -0.137 & 0.027 & -0.009 & -0.077 \\
8 & \hhco\ (3$_{2,2}-2_{2,1}$)  & 0.255 & -0.208 & 0.123 & -0.129 & 0.106 & -0.195 & -0.069 \\
9 & \soo\ (14$_{0,14}-13_{1,13}$)  & 0.255 & 0.112 & -0.124 & 0.176 & -0.081 & 0.046 & -0.098 \\
10 & \mecn\ (12$_{2,0}-11_{2,0}$)  & 0.247 & 0.031 & 0.098 & 0.168 & -0.081 & -0.032 & -0.13 \\
11 & \hhco\ (3$_{2,1}-2_{2,0}$)  & 0.239 & -0.124 & 0.008 & -0.144 & 0.218 & -0.01 & -0.067 \\
12 & CS\ (5$-$4)  & 0.236 & -0.065 & -0.124 & -0.057 & 0.267 & -0.039 & -0.141 \\
13 & \meta\ (18$_{3,16}-17_{4,13}$, A)  & 0.2 & 0.32 & -0.072 & -0.047 & 0.193 & 0.075 & 0.07 \\
14 & HNCO\ (10$_{0,10}-9_{0,9}$)  & 0.185 & 0.26 & 0.033 & 0.176 & 0.239 & -0.077 & 0.318 \\
15 & HNCO\ (12$_{0,12}-11_{0,11}$)  & 0.171 & 0.312 & -0.186 & 0.014 & 0.107 & -0.001 & -0.038 \\
16 & \hhcs\ (7$_{1,7}-6_{1,6}$)  & 0.165 & 0.073 & -0.142 & 0.241 & -0.443 & 0.323 & 0.299 \\
17 & \cchcn\ (11$_{4,8}-10_{3,7}$)  & 0.123 & 0.052 & 0.443 & -0.315 & -0.27 & -0.421 & 0.539 \\
18 & \cchoh\ (12$_{1,11}-11_{2,9}$)  & 0.092 & 0.125 & -0.112 & -0.531 & -0.156 & -0.013 & -0.094 \\
19 & \nhhcho\ (12$_{1,11}-11_{1,10}$)  & 0.078 & 0.351 & -0.01 & 0.391 & 0.1 & -0.284 & 0.163 \\
20 & \hcccn\ (26$-$25)  & 0.07 & -0.274 & 0.241 & 0.409 & -0.41 & -0.223 & -0.287 \\
21 & \nhhcho\ (12$_{0,12}-11_{0,11}$)  & 0.021 & 0.454 & -0.173 & -0.126 & -0.325 & -0.37 & -0.419 \\
22 & \co\ (2$-$1)  & -0.055 & -0.373 & -0.603 & 0.106 & 0.103 & -0.556 & 0.211 \\
23 & SiO\ (6$-$5)  & -0.062 & 0.044 & 0.434 & 0.213 & 0.388 & -0.113 & -0.163 \\\hline
\hline
&Eigen values&10.86  & 2.758 & 1.62  & 1.439  &1.198 & 1.051 & 0.916   \\
&Contribution ratio (\%)&45.0& 11.4&  6.7&  6.0&   5.0&   4.4&  3.8 \\ 
\hline
\end{tabular}
}
\begin{flushleft}
\tablecomments{These values are also called as 'loadings'.
}
\end{flushleft}
\end{table}

\begin{table}[ht]
\caption{Eigenvectors of the Principal Components and their Eigenvalues for PCA-2D \label{percent_2d}}
\scalebox{1.0}{
\begin{tabular}{ccccccccccccc}
\hline \hline
Number & Molecule & PC1 & PC2 & PC3 & PC4 & PC5 & PC6 & PC7 \\
\hline
1 & \meta\ (5$_{1,4}-4_{1,3}$, A) & 0.224 & 0.115 & 0.101 & 0.171 & -0.159 & -0.328 & -0.383 \\
2 & \meta\ (4$_{2,2}-5_{1,5}$, A)  & 0.222 & 0.031 & 0.031 & 0.103 & -0.228 & -0.252 & -0.201 \\
3 & SO\ (5$_6-4_5$) & 0.223 & 0.175 & -0.075 & 0.214 & 0.013 & 0.269 & 0.325 \\
4 & \meta\ (4$_{2,3}-3_{1,2}$, E)  & 0.224 & 0.068 & 0.051 & 0.092 & 0.197 & -0.046 & -0.179 \\
5 & SO\ (6$_7-5_6$)  & 0.212 & 0.231 & -0.006 & 0.202 & -0.29 & 0.212 & 0.289 \\
6 & \meta\ (10$_{3,7}-11_{2,9}$, E)  & 0.223 & -0.018 & -0.023 & 0.002 & 0.226 & 0.032 & 0.01 \\
7 & \hhco\ (3$_{2,1}-2_{2,0}$)  & 0.224 & 0.107 & 0.015 & 0.152 & 0.182 & -0.138 & -0.212 \\
8 & \hhco\ (3$_{2,2}-2_{2,1}$)  & 0.224 & 0.116 & -0.011 & 0.188 & 0.143 & -0.06 & -0.071 \\
9 & \soo\ (14$_{0,14}-13_{1,13}$)  & 0.219 & 0.023 & 0.085 & -0.036 & -0.287 & 0.269 & 0.092 \\
10 & \mecn\ (12$_{2,0}-11_{2,0}$)  & 0.225 & -0.025 & 0.096 & -0.061 & -0.035 & 0.139 & -0.14 \\
11 & \hhco\ (3$_{2,1}-2_{2,0}$)  & 0.219 & 0.152 & -0.068 & 0.176 & 0.192 & -0.052 & -0.016 \\
12 & CS\ (5$-$4)  & 0.218 & 0.23 & -0.11 & 0.268 & -0.088 & -0.029 & 0.222 \\
13 & \meta\ (18$_{3,16}-17_{4,13}$, A)  & 0.217 & -0.132 & -0.133 & -0.127 & 0.267 & -0.049 & 0.108 \\
14 & HNCO\ (10$_{0,10}-9_{0,9}$)  & 0.214 & -0.079 & 0.014 & -0.114 & 0.373 & 0.075 & -0.093 \\
15 & HNCO\ (12$_{0,12}-11_{0,11}$)  & 0.224 & -0.12 & -0.102 & -0.165 & -0.123 & 0.057 & 0.165 \\
16 & \hhcs\ (7$_{1,7}-6_{1,6}$)  & 0.222 & -0.12 & 0.112 & -0.276 & 0.099 & 0.369 & 0.005 \\
17 & \cchcn\ (11$_{4,8}-10_{3,7}$)  & 0.201 & -0.185 & -0.133 & -0.309 & -0.473 & -0.32 & 0.068 \\
18 & \cchoh\ (12$_{1,11}-11_{2,9}$)  & 0.223 & -0.066 & -0.114 & -0.04 & -0.082 & -0.105 & 0.107 \\
19 & \nhhcho\ (12$_{1,11}-11_{1,10}$)  & 0.199 & -0.197 & -0.269 & -0.154 & 0.075 & -0.341 & 0.185 \\
20 & \hcccn\ (26$-$25)  & 0.201 & 0.019 & 0.4 & -0.211 & -0.218 & 0.268 & -0.383 \\
21 & \nhhcho\ (12$_{0,12}-11_{0,11}$)  & 0.21 & -0.21 & -0.089 & -0.244 & 0.147 & 0.075 & 0.003 \\
22 & \co\ (2$-$1)  & -0.03 & 0.766 & 0.067 & -0.58 & 0.107 & -0.187 & 0.117 \\
23 & SiO\ (6$-$5)  & 0.063 & -0.182 & 0.794 & 0.07 & 0.12 & -0.319 & 0.454 \\
\hline
\hline
&Eigen values& 19.37 &  2.011&  1.377 & 0.602&  0.239 & 0.135 & 0.087 \\
&Contribution ratio (\%) & 77.9 & 8.1  &5.5 & 2.4 & 1.0  & 0.5&  0.3   \\
\hline
\end{tabular}
}
\begin{flushleft}
\tablecomments{These values are also called as 'loadings'.
}
\end{flushleft}
\end{table}

\begin{table}[ht]
\caption{Eigenvectors of the Principal Components and their Eigenvalues for PCA-1D \label{percent_1d}}
\scalebox{1.0}{
\begin{tabular}{ccccccccccccc}
\hline \hline
Number & Molecule & PC1 & PC2 & PC3 & PC4 & PC5 & PC6 & PC7 \\
\hline
1 & \meta\ (5$_{1,4}-4_{1,3}$, A) & 0.285 & 0.109 & 0.057 & 0.054 & -0.107 & 0.03 & -0.017 \\
2 & \meta\ (4$_{2,2}-5_{1,5}$, A)  & 0.28 & 0.039 & 0.091 & 0.066 & -0.011 & -0.071 & 0.027 \\
3 & SO\ (5$_6-4_5$) & 0.276 & 0.14 & -0.013 & 0.053 & -0.072 & -0.146 & -0.289 \\
4 & \meta\ (4$_{2,3}-3_{1,2}$, E)  & 0.281 & 0.106 & -0.006 & 0.057 & 0.005 & -0.147 & 0.183 \\
5 & SO\ (6$_7-5_6$)  & 0.283 & 0.105 & 0.002 & 0.01 & -0.079 & -0.034 & -0.252 \\
6 & \meta\ (10$_{3,7}-11_{2,9}$, E)  & 0.27 & 0.093 & -0.063 & 0.211 & 0.086 & -0.165 & -0.112 \\
7 & \hhco\ (3$_{2,1}-2_{2,0}$)  & 0.263 & 0.208 & 0.032 & -0.008 & 0.053 & -0.044 & 0.122 \\
8 & \hhco\ (3$_{2,2}-2_{2,1}$)  & 0.267 & 0.172 & 0.1 & 0.051 & 0.149 & -0.255 & -0.092 \\
9 & \soo\ (14$_{0,14}-13_{1,13}$)  & 0.254 & -0.201 & -0.088 & -0.245 & -0.184 & 0.014 & -0.418 \\
10 & \mecn\ (12$_{2,0}-11_{2,0}$)  & 0.238 & 0.158 & 0.007 & -0.068 & 0.066 & 0.726 & 0.307 \\
11 & \hhco\ (3$_{2,1}-2_{2,0}$)  & 0.272 & -0.008 & -0.006 & 0.214 & 0.179 & 0.231 & 0.094 \\
12 & CS\ (5$-$4)  & 0.266 & -0.148 & 0.02 & 0.133 & 0.186 & 0.074 & 0.002 \\
13 & \meta\ (18$_{3,16}-17_{4,13}$, A)  & 0.191 & -0.361 & -0.18 & -0.076 & -0.211 & -0.428 & 0.565 \\
14 & HNCO\ (10$_{0,10}-9_{0,9}$)  & 0.184 & -0.33 & -0.212 & 0.05 & -0.658 & 0.273 & -0.047 \\
15 & HNCO\ (12$_{0,12}-11_{0,11}$)  & 0.106 & -0.502 & -0.123 & -0.328 & 0.453 & 0.091 & -0.306 \\
16 & \hhcs\ (7$_{1,7}-6_{1,6}$)  & 0.146 & 0.272 & -0.008 & -0.824 & 0.002 & -0.057 & 0.175 \\
17 & \cchoh\ (12$_{1,11}-11_{2,9}$)  & -0.009 & -0.075 & 0.853 & -0.111 & -0.291 & 0.013 & -0.087 \\
18 & \hcccn\ (26$-$25)  & -0.152 & 0.453 & -0.38 & -0.06 & -0.268 & 0.013 & -0.225 \\
\hline
\hline
&Eigen values&11.77 & 2.564  &1.31  & 0.755&  0.426&  0.348 & 0.318\\
&Contribution ratio (\%) &64.1 &14.0   &7.1 & 4.1 & 2.3 & 1.9&  1.7 \\
\hline
\end{tabular}
}
\begin{flushleft}
\tablecomments{These values are also called as 'loadings'.
}
\end{flushleft}
\end{table}



\begin{figure}[h!]
\centering
\includegraphics[scale=0.4]{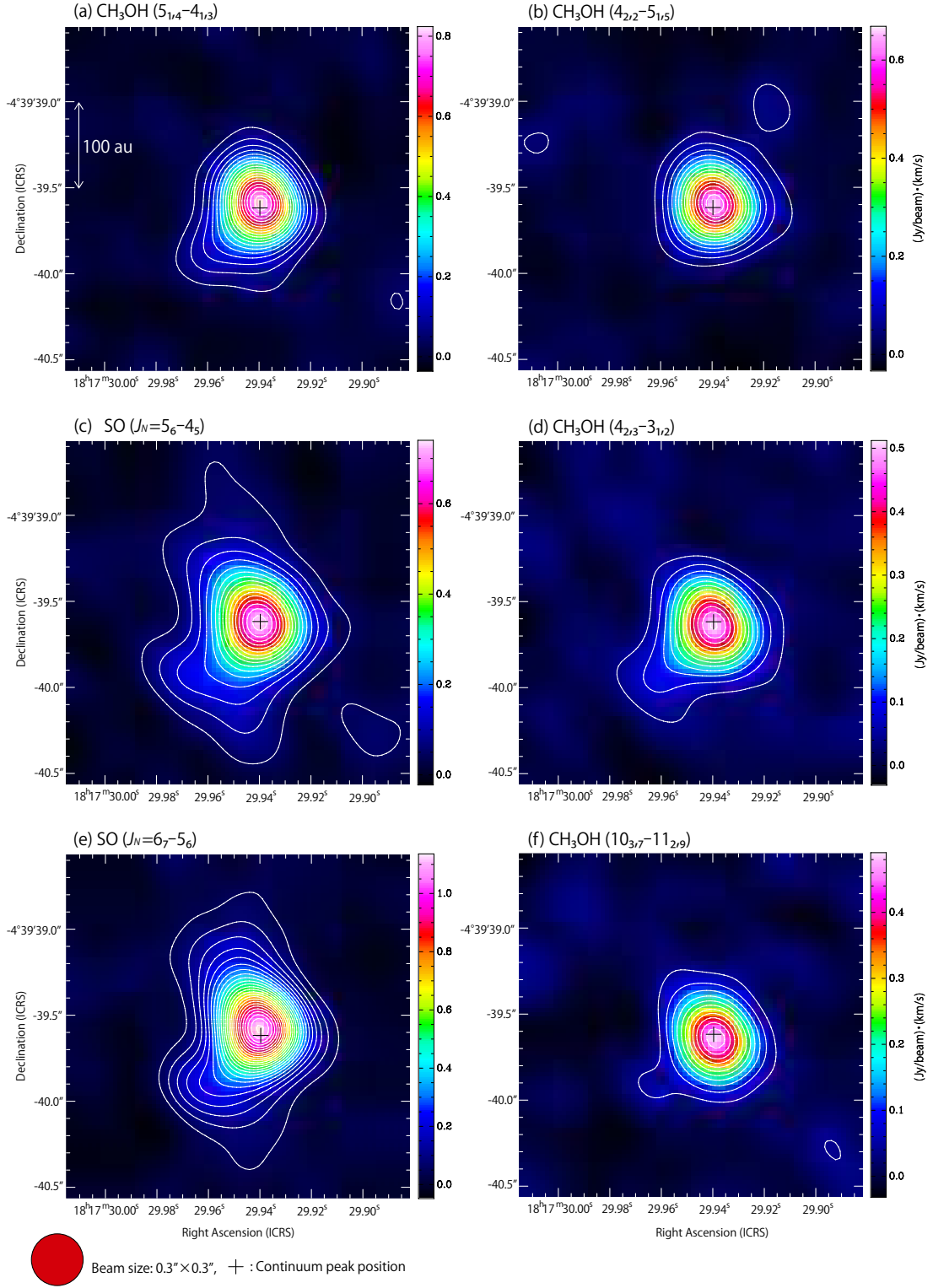}
\caption{(a-w) Moment 0 maps of the 23 molecular lines. The cross marks show the continuum peak position: ($\alpha_{2000}$, $\delta_{2000}$)=(18\fh17\fm29\fs940, $-$04\arcdeg 39\arcmin39\farcs60).
 The order of (a)-(w) is the same as that in Table \ref{observations} (1-23).
The integrated velocity range is from -2.9 \kms\ to 14.25 \kms. 
Contour levels are every 3$\sigma$ from 3$\sigma$, where $\sigma$ is listed in Table \ref{observations}. \label{moment_mol}}
\end{figure}
\addtocounter{figure}{-1}
\begin{figure}[h!]
\centering
\includegraphics[scale=0.4]{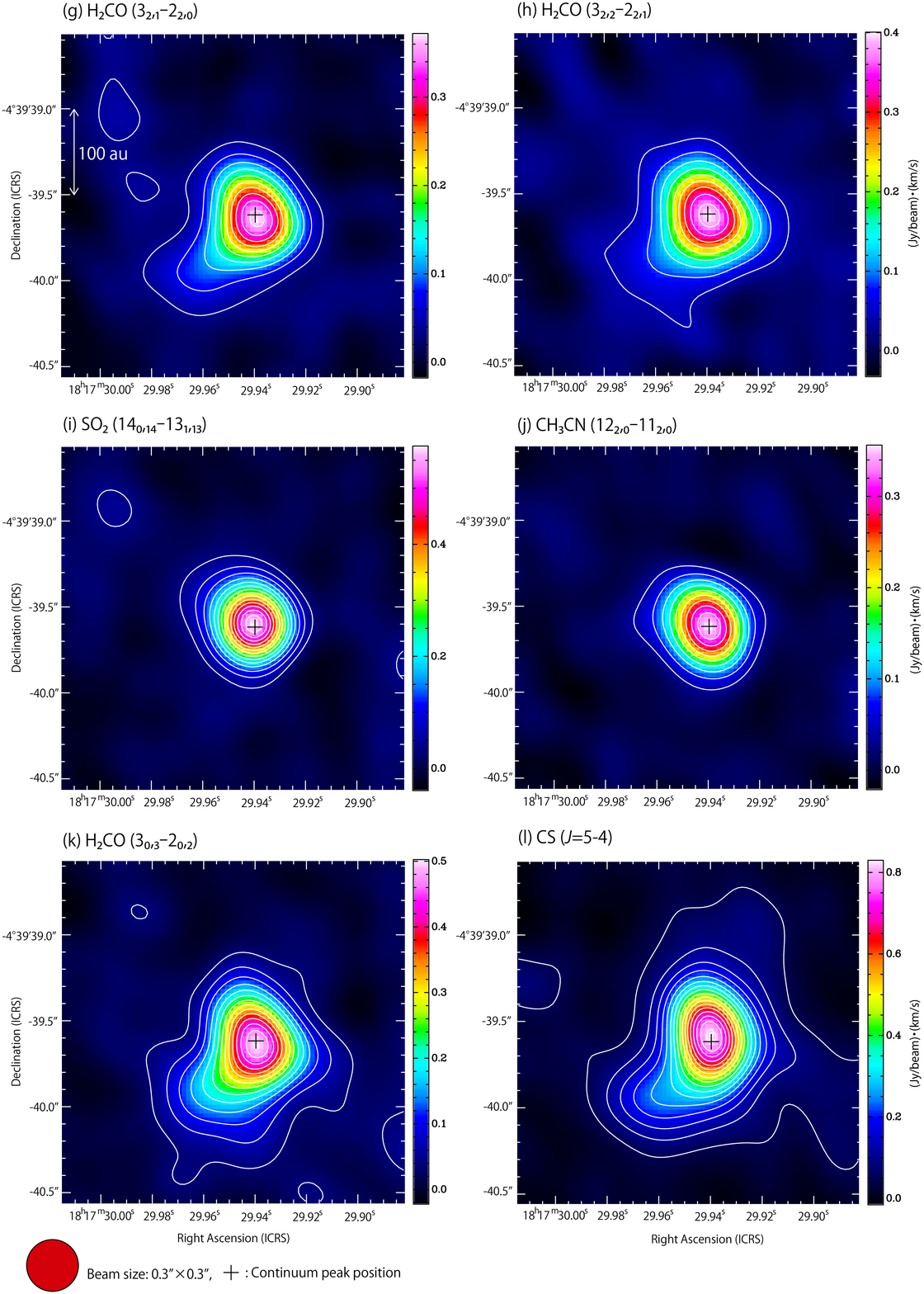}
\caption{(Continued.)\label{moment_mol}}
\end{figure}
\addtocounter{figure}{-1}
\begin{figure}[h!]
\centering
\includegraphics[scale=0.4]{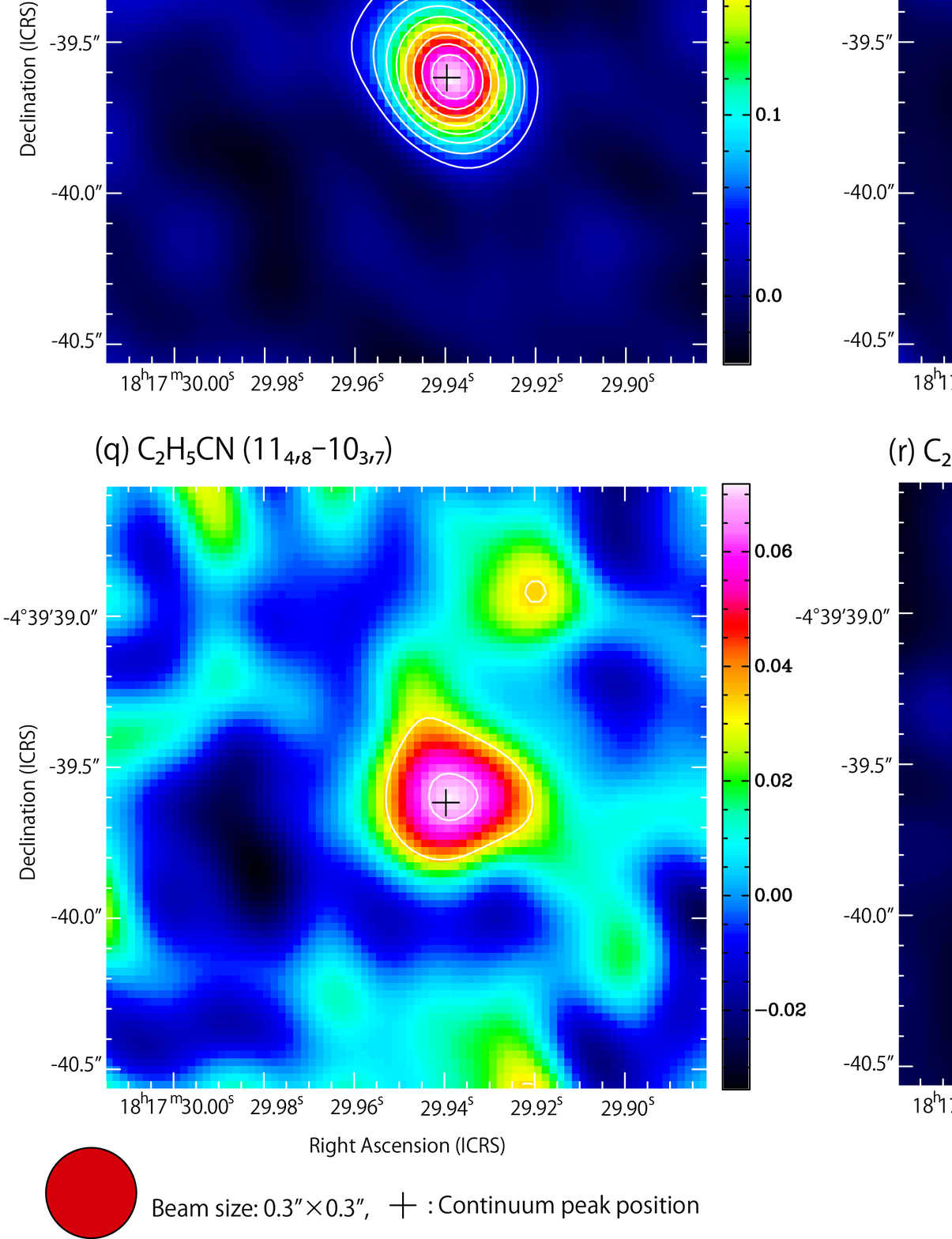}
\caption{(Continued.)\label{moment_mol}}
\end{figure}
\addtocounter{figure}{-1}
\begin{figure}[h!]
\centering
\includegraphics[scale=0.4]{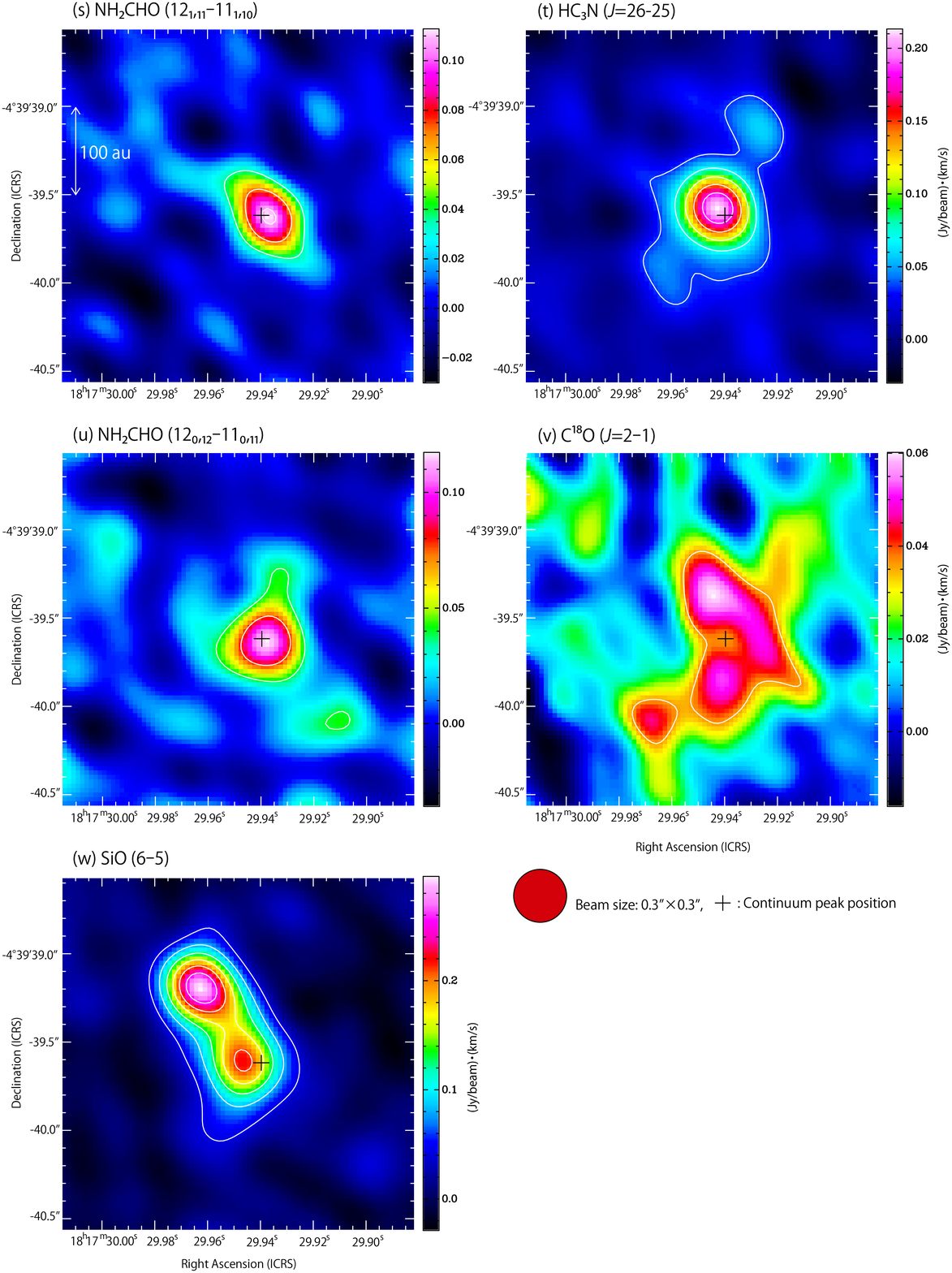}
\caption{(Continued.)\label{moment_mol}}
\end{figure}

\begin{figure}[h!]
\centering
\includegraphics[scale=0.35]{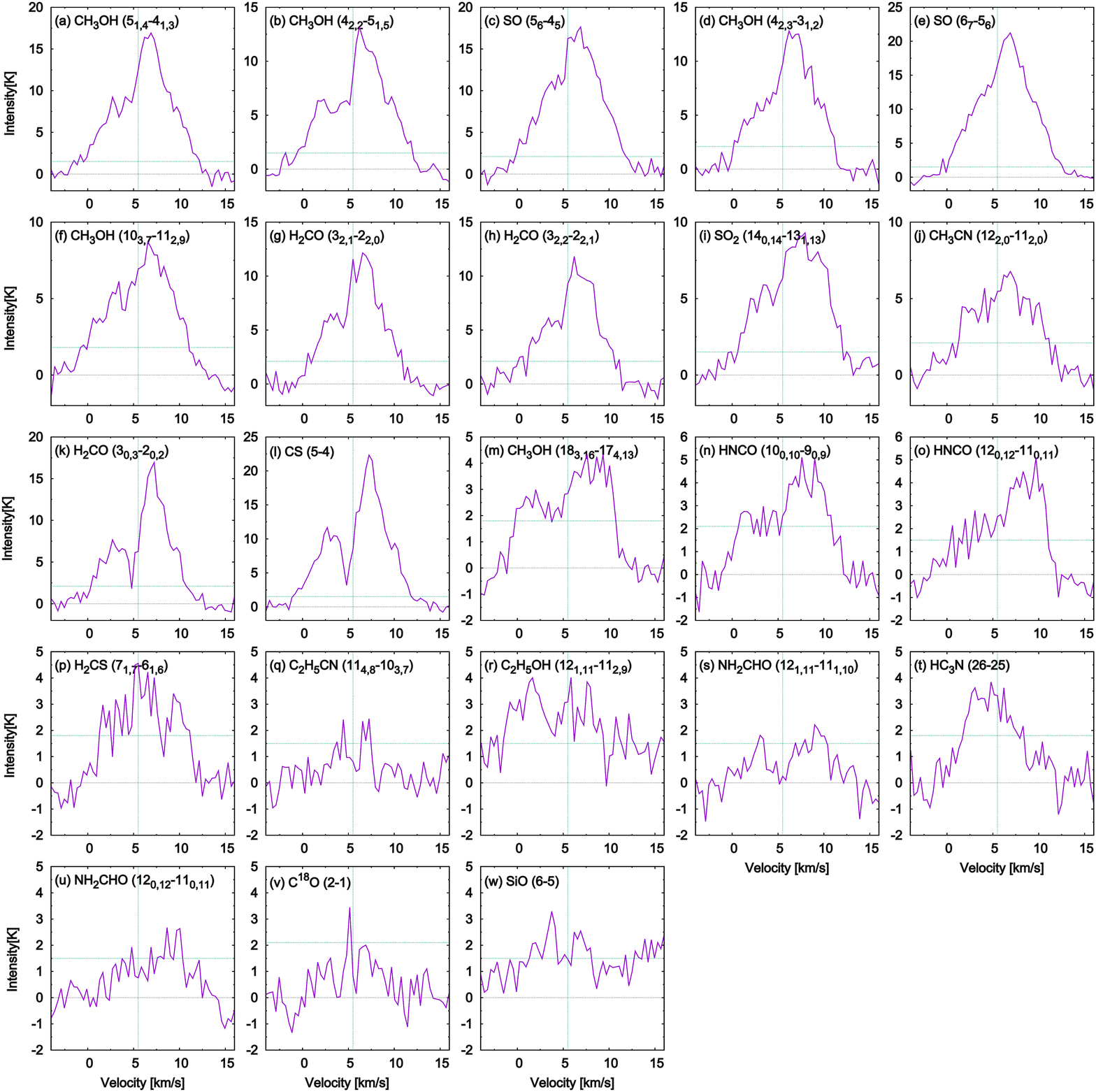}
\caption{(a-w) Molecular line profiles observed toward the continuum peak.
The spectra are prepared for a circular region with a diameter of 0\farcs5 centered at the continuum peak.
The order of (a)-(w) is the same as that in Table \ref{observations} (1-23).
The horizontal green dashed line in each panel represents 3$\sigma$ for each spectrum, where $\sigma$ is listed in Table \ref{observations}. The horizontal black dashed line in each panel represents the zero-level intensity. The vertical green dashed lines represent the systemic velocity of 5.5 \kms\ \citep{Hirota et al.(2009)}. \label{spectral_mol}}
\end{figure}

\begin{figure}[h!]
\rotatebox{90}{
\begin{minipage}{\textheight}
\centering
\includegraphics[scale=0.5]{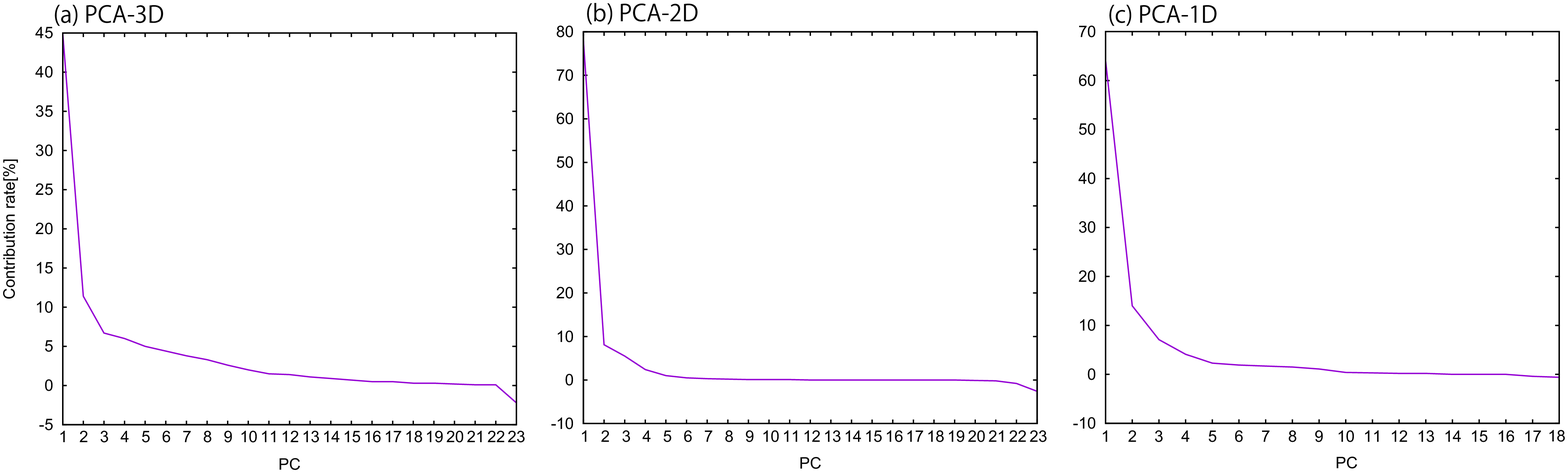}
\caption{Scree plots of the contribution ratios for the principal components. (a) PCA for the cube data (Section \ref{sec-3d}; Table \ref{percent_3d}). (b) PCA for the moment 0 maps (Section \ref{sec-2d}; Table \ref{percent_2d}). (c) PCA for the spectral line profiles (Section \ref{sec-1d}; Table \ref{percent_1d}). Negative contribution ratios found in higher-order PCs are due to the observation noise.\label{screeplot}}
\end{minipage}}
\end{figure}


\begin{figure}[h!]
\rotatebox{90}{
\begin{minipage}{\textheight}
\centering
\includegraphics[scale=0.3]{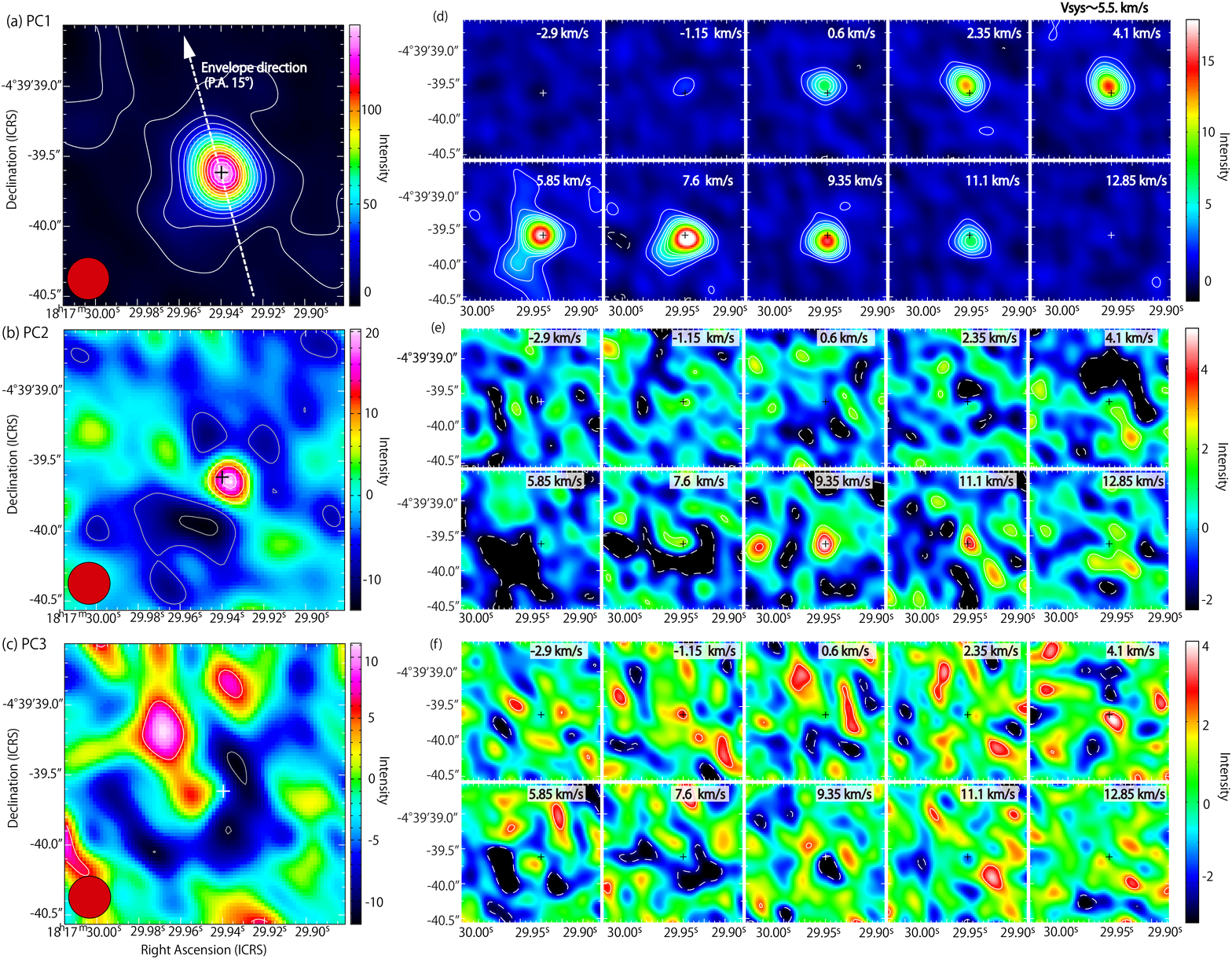}
\caption{(a-c) Moment 0 maps of the first three principal components in PCA-3D. 
Contour levels for PC1 (a) is every 10$\sigma$ from 3$\sigma$, where $\sigma$ is 1.
Those for PC2 (b) and PC3 (c) are every 3$\sigma$ from 3$\sigma$ where $\sigma$ is 2.
The red circles show the beam size.
(d-f) Channel maps of the first three principal components. Each panel represents the integrated intensity over a velocity range of 1.75 \kms, whose lower-end velocity is quoted on the upper right corner. The systemic velocity is 5.5 \kms. Contour levels for (d) PC1 are 3$\sigma$, 6$\sigma$, and 12 $\sigma$, where $\sigma$ is 0.4.
Contour levels for (e) PC2 and (f) PC3 are every 3$\sigma$ from 3$\sigma$, where $\sigma$ is 0.6 and 0.8, respectively.
The cross marks show the continuum peak position. \label{result3d}}
\end{minipage}}
\end{figure}

\begin{figure}[h!]
\centering
\includegraphics[scale=0.8]{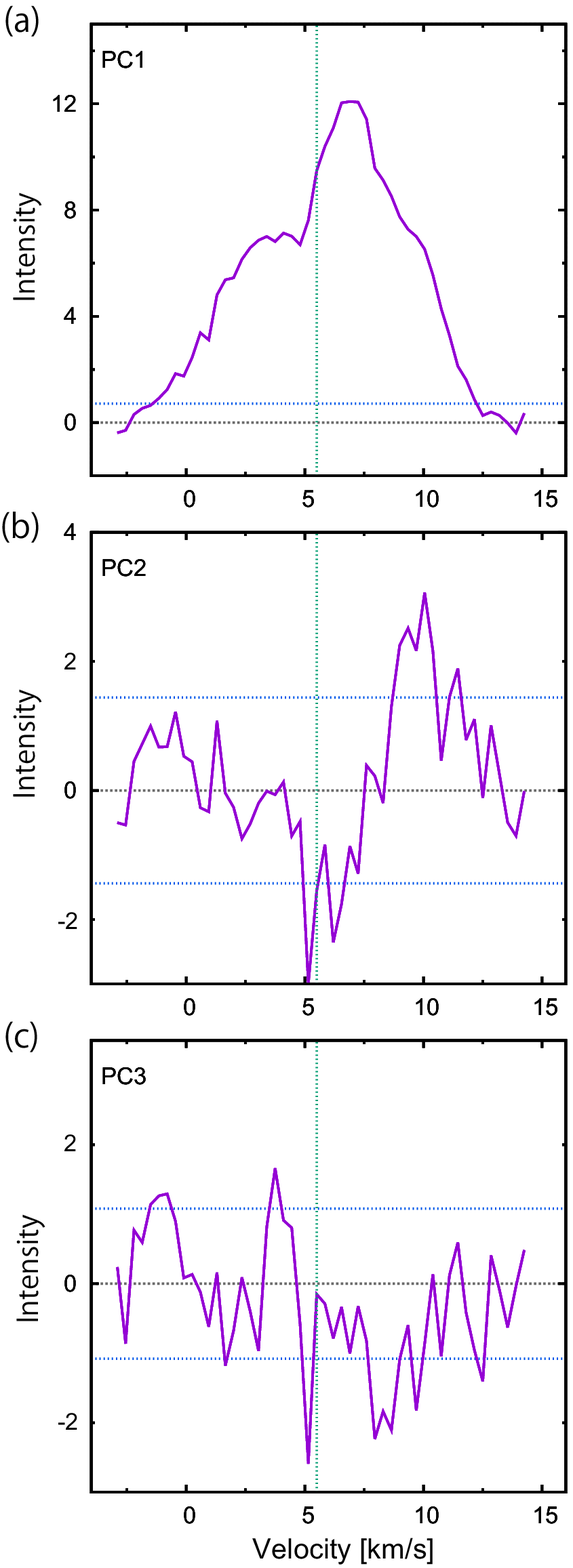}
\caption{Spectral line profiles of the first three principal components in PCA-3D toward the  continuum peak. 
The spectra are prepared for a circular region with a diameter of 0\farcs5 centered at the continuum peak.
The horizontal blue dashed lines represent each $\pm$3$\sigma$, where $\sigma$ of (a) PC1, (b) PC2, and (c) PC3 are 0.24, 0.36, and 0.48, respectively. The horizontal black dashed lines represent the zero-level intensity. The vertical green dashed lines represent the systemic velocity of 5.5 \kms. 
 \label{spectral_pc}}
\end{figure}

\begin{figure}[h!]
\centering
\includegraphics[scale=0.5]{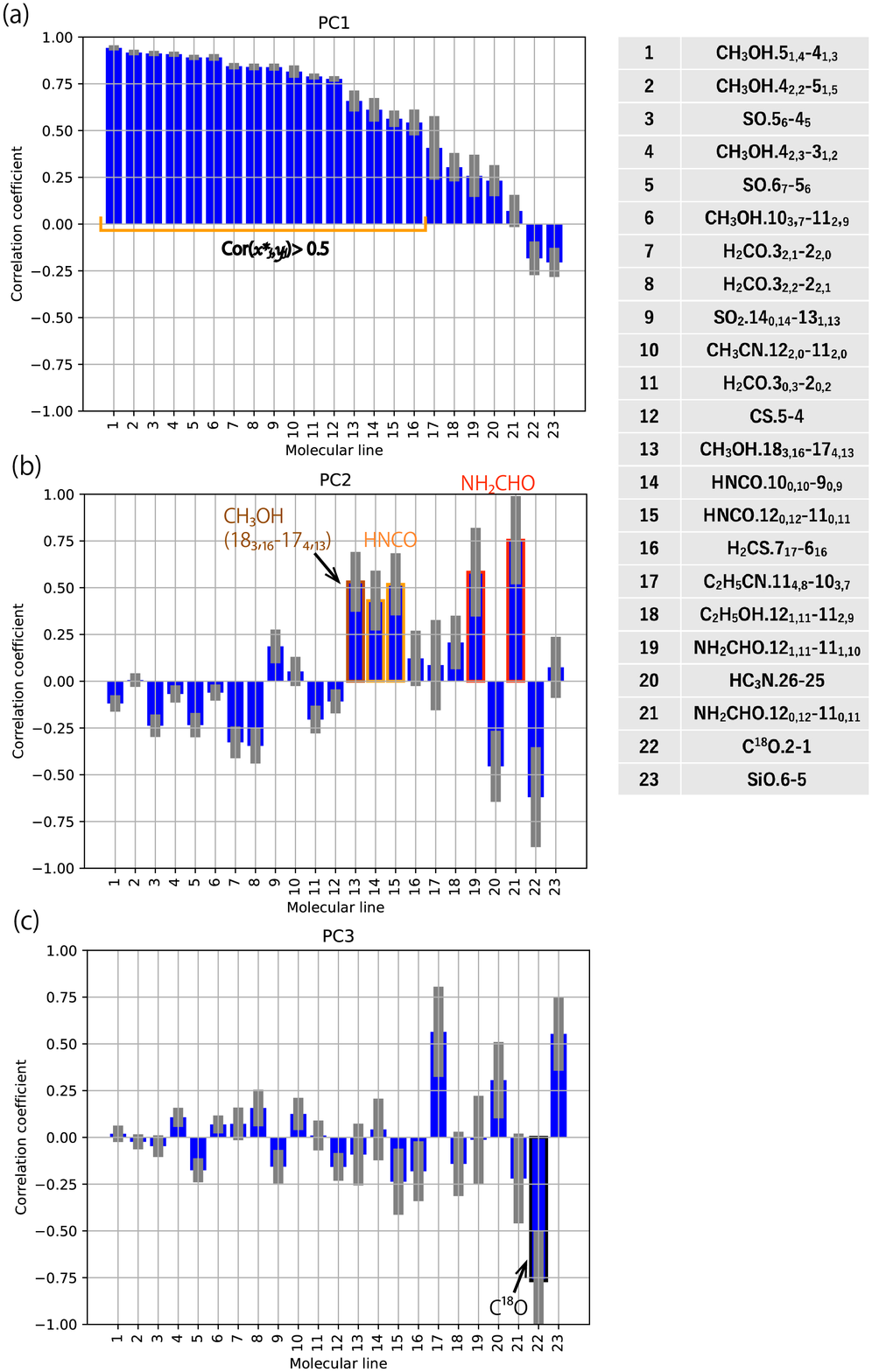}
\caption{Correlation coefficients between the first three principal components in PCA-3D and the molecular cube data. The numbers represent the molecular lines listed in the attached table. The uncertainties are shown in grey (See Appendix \ref{SD_PCA}).  \label{loading_3d}}
\end{figure}

\begin{figure}[h!]
\rotatebox{90}{
\begin{minipage}{\textheight}
\centering
\includegraphics[scale=0.35]{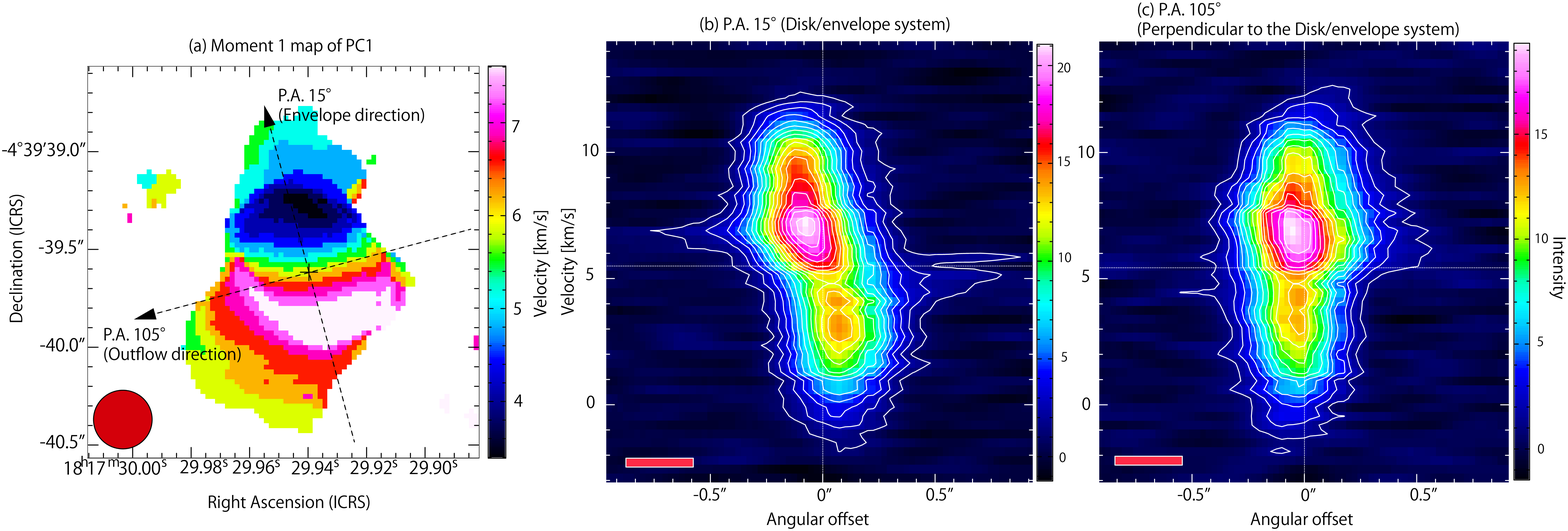}
\caption{(a) Moment 1 map of PC1 in PCA-3D. The red circle shows the beam size. The cross marks show the continuum peak position. (b, c) PV diagrams of PC1 along the disk/envelope direction indicated in (a) and the direction perpendicular to it, respectively. Contour levels are every 3$\sigma$ from 3$\sigma$, where $\sigma$ is 0.4. The horizontal and vertical white dashed lines show the systemic velocity (5.5 \kms) and the continuum peak position, respectively. The red boxes represent the resolution. \label{moment1_pc1}}
\end{minipage}}
\end{figure}

\begin{figure}[h!]
\centering
\includegraphics[scale=0.4]{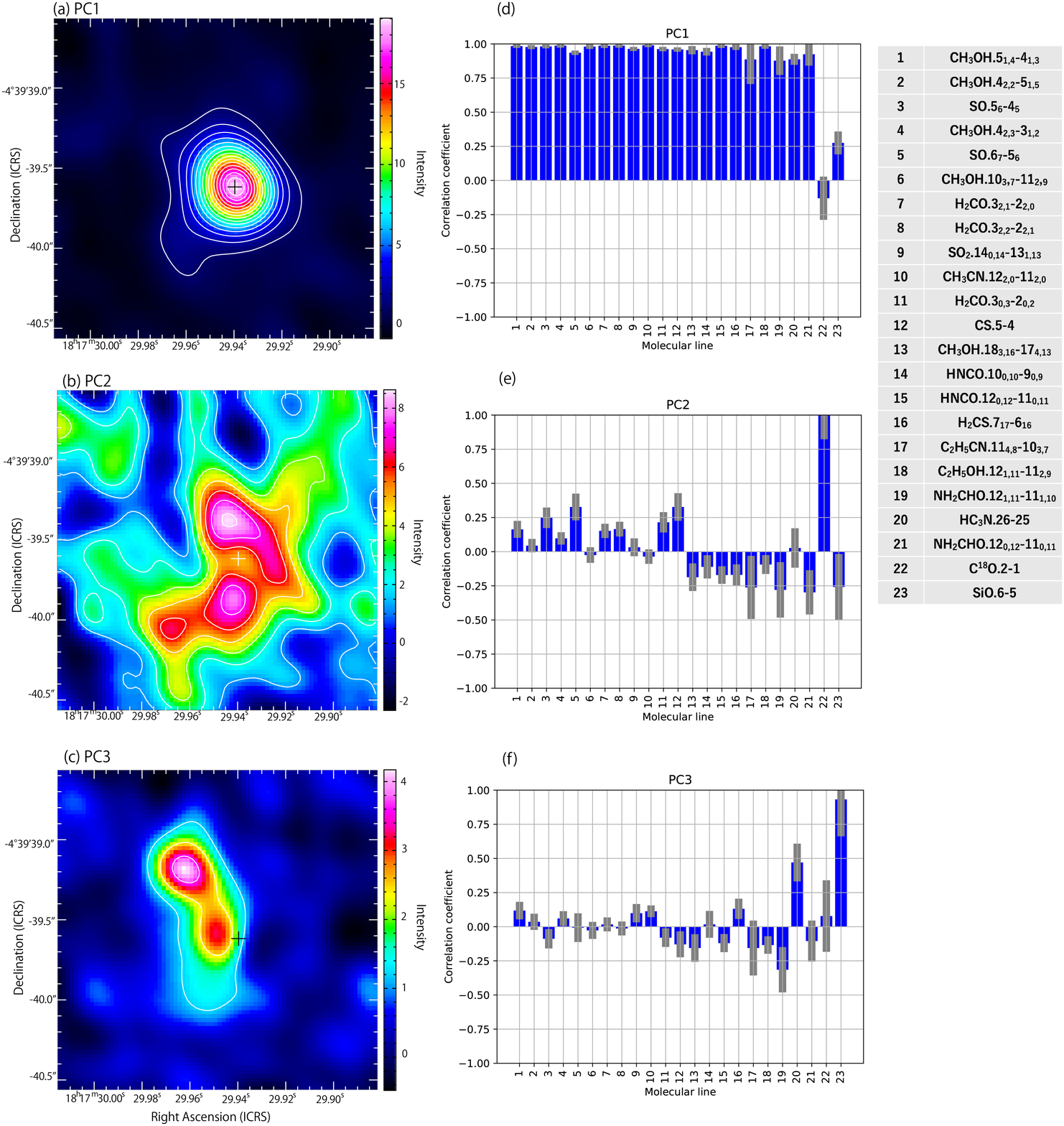}
\caption{(a-c) Moment 0 maps of the first three principal components in PCA-2D.  Contour levels for the first three components are every 3$\sigma$ from 3$\sigma$, where $\sigma$ is 0.4. (d-f) Correlation coefficients between the first three principal components in PCA-2D and the molecular-line distributions. The uncertainties are shown in grey (See Appendix \ref{SD_PCA}). The numbers represent the molecular lines listed in the attached table.\label{result_2d}}
\end{figure}

\begin{figure}[h!]
\centering
\includegraphics[scale=0.55]{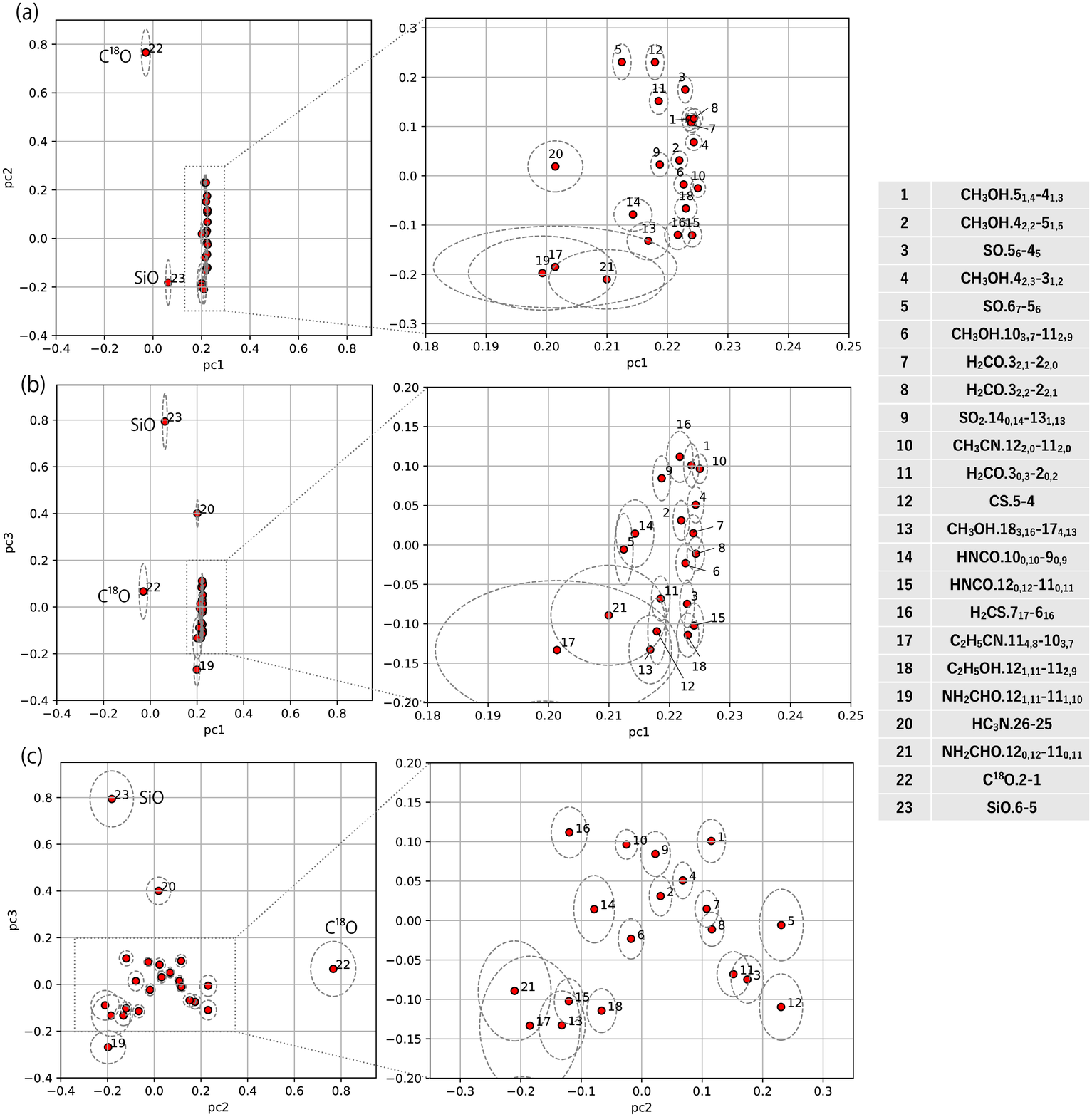}
\caption{(a-c)  Biplots of the contributions for the principal components in PCA-2D for each molecular-line distribution on the PC1-PC2, PC1-PC3, and PC2-PC3 planes. The dashed ellipses represent the uncertainties (See Appendix \ref{SD_PCA}). The numbers represent the molecular lines listed in the attached table.\label{result_2d_biplot}}
\end{figure}

\begin{figure}[h!]
\centering
\includegraphics[scale=0.5]{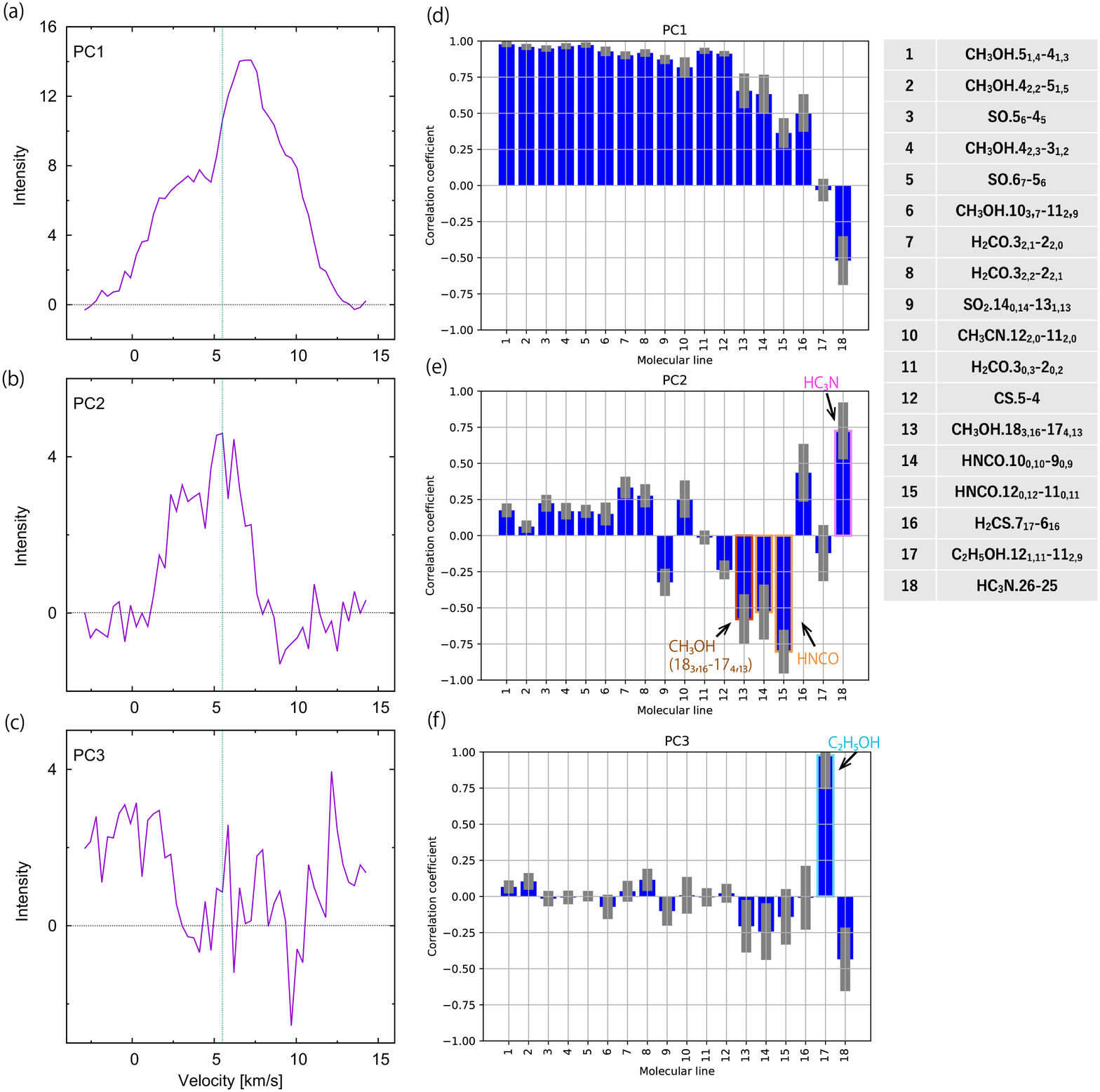}
\caption{(a-c) Spectral line profiles of the first three principal components in PCA-1D toward the continuum peak position.
The spectra are prepared for a circular region with a diameter of 0\farcs5 centered at the continuum peak.
The horizontal black dashed lines represent the zero-level intensity. 
The vertical green dashed lines represent the systemic velocity of 5.5 \kms. 
(d-f) Correlation coefficients between the first three principal components in PCA-1D and the molecular-line profiles. The uncertainties are shown in grey (See Appendix \ref{SD_PCA}). The numbers represent the molecular lines listed in the attached table.\label{result_1d_a}}
\end{figure}

\begin{figure}[h!]
\centering
\includegraphics[scale=0.55]{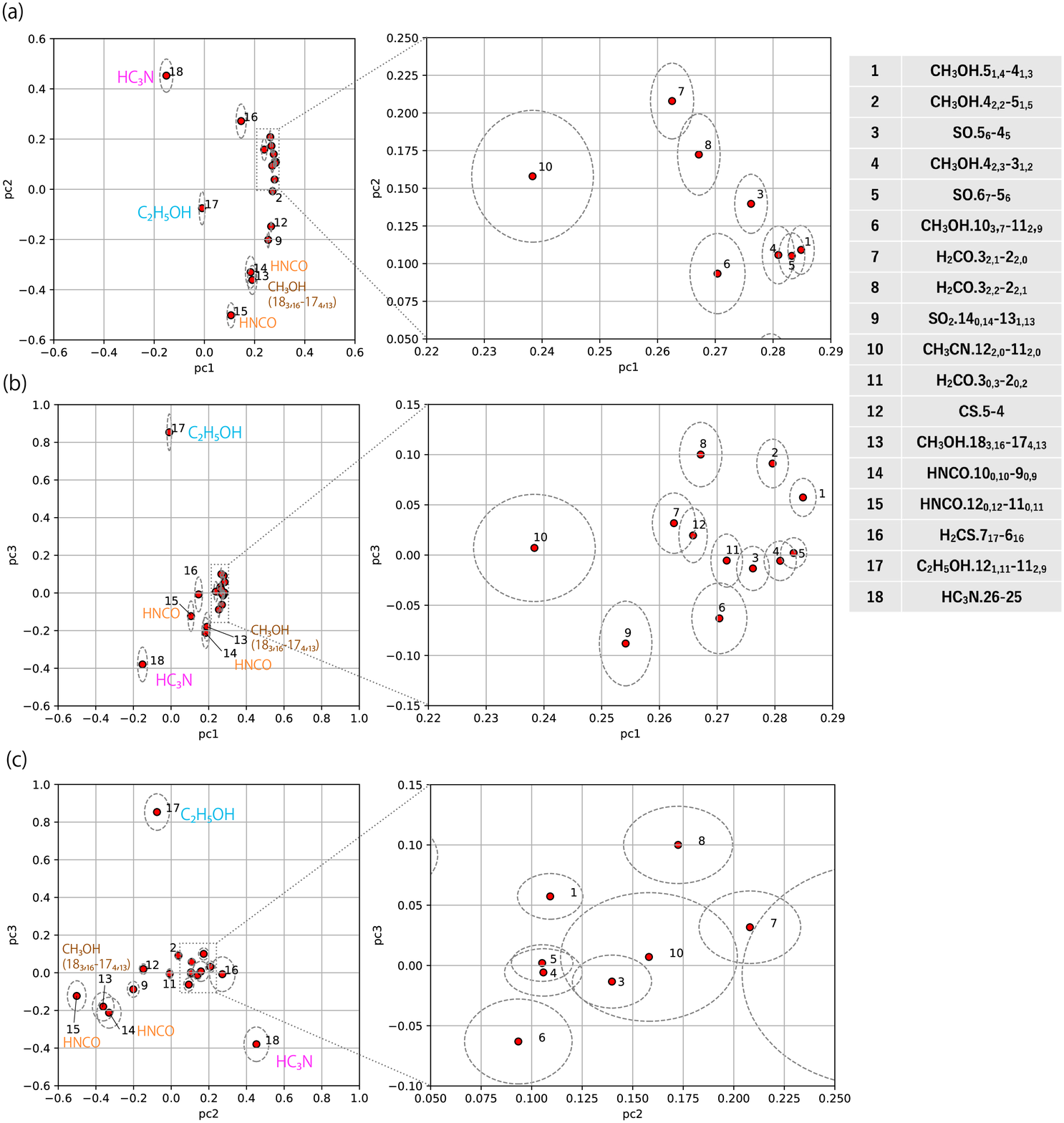}
\caption{(a-c) Biplots of the contributions for the principal components in PCA-1D for each molecular-line distribution on the PC1-PC2, PC1-PC3, and PC2-PC3 planes. The dashed ellipses represent the uncertainties (See Appendix \ref{SD_PCA}). The numbers represent the molecular lines listed in the attached table.\label{result_1d_b}}
\end{figure}

\begin{figure}[h!]
\centering
\includegraphics[scale=0.55]{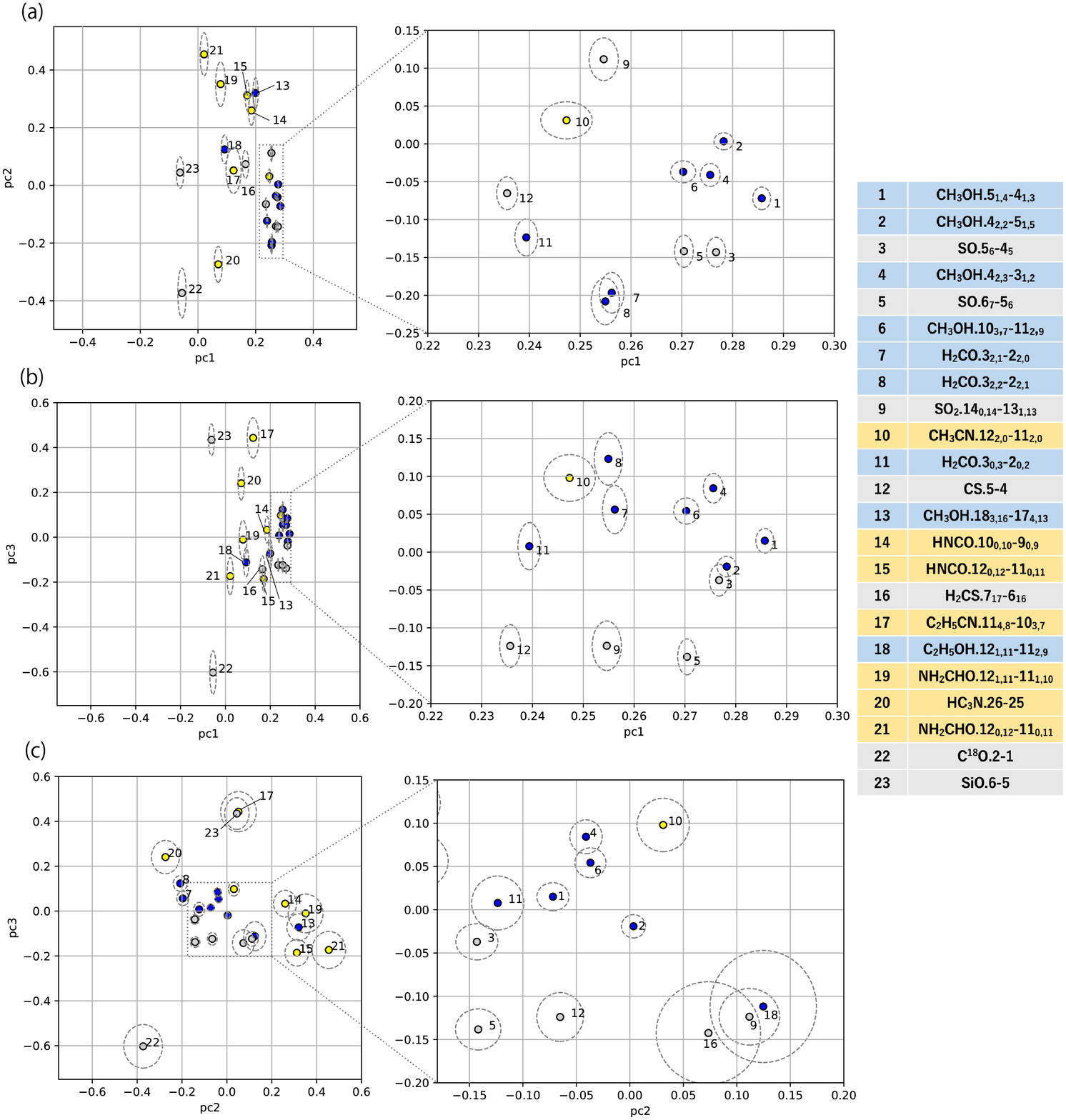}
\caption{(a-c) Biplots of the contributions for the principal components in PCA-3D for each molecular-line distribution on the PC1-PC2, PC1-PC3, and PC2-PC3 planes. The dashed ellipses represent the uncertainties (See Appendix \ref{SD_PCA}). The numbers represent the molecular lines listed in the attached table.
Blue and yellow represent the oxygen-bearing and nitrogen-bearing molecules, respectively. \label{plot_3d}}
\end{figure}

\end{document}